\documentclass[]{aa}
\usepackage{graphicx}
\usepackage{psfig}
\usepackage{rotating}
\def\h2o{${\rm H_2O}$}
\def\pht{\rm ISO-PHT-S~}
\def\cam{\rm ISO-CAM-CVF~}
\def\mum{${\mu}\rm m$}
\def\ltsim{${_<\atop{^\sim}}$}

\begin{document}

\title{Ice features in the mid-IR spectra of galactic nuclei
\thanks{Based on observations with ISO, an ESA project with 
       instruments funded by ESA Member States (especially the PI 
       countries: France, Germany, the Netherlands and the United 
       Kingdom) and with the participation of ISAS and NASA}}
\author{H.W.W. Spoon\inst{1,2} 
   \and J.V. Keane\inst{2}
   \and A.G.G.M. Tielens\inst{2,3}
   \and D. Lutz\inst{4}
   \and A.F.M. Moorwood\inst{1}
   \and O. Laurent\inst{4}}

\offprints{H.W.W. Spoon}

\institute{European Southern Observatory, Karl-Schwarzschild-Strasse 2,
           D-85748 Garching, Germany\\
           email: amoor@eso.org
        \and
           Kapteyn Institute, P.O. Box 800, NL-9700 AV Groningen, 
           the Netherlands\\
           email: spoon@astro.rug.nl, j.keane@astro.rug.nl, 
           tielens@astro.rug.nl
       \and
           SRON, P.O. Box 800, NL-9700 AV Groningen, the Netherlands
       \and
           Max-Planck-Institut f\"ur Extraterrestrische Physik (MPE),
           Postfach 1312, 85741 Garching, Germany\\
           email: lutz@mpe.mpg.de}
\date{Received date; accepted date}

\abstract{Mid infrared spectra provide a powerful probe of the conditions 
in dusty galactic nuclei. They variously contain emission features 
associated with star forming regions and absorptions by circumnuclear 
silicate dust plus ices in cold molecular cloud material. Here we report 
the detection of 6--8\,$\mu$m water ice absorption
in 18 galaxies observed by ISO. While the mid-IR spectra of some of 
these galaxies show a strong resemblance to the heavily absorbed 
spectrum of NGC\,4418, other galaxies in this sample also show weak 
to strong PAH emission. The 18 ice galaxies are part of a sample of 
103 galaxies with good S/N mid-IR ISO spectra. Based on our sample we 
find that ice is present in most of the ULIRGs, whereas it is weak 
or absent in the large majority of Seyferts and starburst galaxies. 
This result is consistent with the presence of larger quantities of 
molecular material in ULIRGs as opposed to other galaxy types.\\
Like NGC\,4418, several of our ice galaxy spectra show a maximum 
near 8\,$\mu$m that is not or only partly due to PAH emission. While 
this affects only a small part of the galaxy population studied by 
ISO, it stresses the need for high S/N data and refined diagnostic 
methods, to properly discriminate spectra dominated by PAH emission 
and spectra dominated by heavy obscuration.\\
The spectral variation from PAH emission to absorbed continuum emission 
near 8\,$\mu$m shows strong similarities with Galactic star forming 
clouds. This leads us to believe that our classification of ice galaxy 
spectra might reflect an evolutionary sequence from strongly obscured 
beginnings of star formation (and AGN activity) to a less enshrouded 
stage of advanced star formation (and AGN activity), as the PAHs get 
stronger and the broad 8\,$\mu$m feature weakens.
\keywords{Galaxies: ISM --- Galaxies: nuclei --- Galaxies: Seyfert --- 
Galaxies: starburst --- Infrared: galaxies --- Infrared: ISM}
}

\maketitle

\section{Introduction}

Prior to the ISO mission, the mid-IR spectra of galaxies could only be
studied from the ground in certain windows of reduced atmospheric 
absorption. For the far brighter Galactic sources less limitations 
applied, as these could be studied with airborne telescopes, like the 
Kuiper Airborne Observatory (KAO: 1974-1995). 
Hence, before the advent of ISO, most of the prominent mid-IR ISM 
features had already been studied in some detail for Galactic sources, 
but not yet for extragalactic sources.
Equipped with three mid-IR spectro(photo)meters, ISO 
has since enhanced considerably our knowledge of the mid-IR spectral
properties of normal, starburst, Seyfert and Ultra-luminous Infrared 
Galaxies (ULIRGs).

The mid-IR spectra of most galaxies are dominated by ISM emission 
features, the most prominent of which are the well-known PAH emission 
bands at 6.2, 7.7, 8.6, 11.3 and 12.7\,$\mu$m and atomic emission lines. 
The PAH features and the emission lines have been used qualitatively
and quantitatively as diagnostics for the ultimate physical processes
powering galactic nuclei (Genzel et al. \cite{Genzel}; Lutz
et al. \cite{Lutz98}; Rigopoulou et al. \cite{Rigopoulou}; Clavel et
al. \cite{Clavel}; Helou et al. \cite{Helou}; Tran et al.
\cite{Tran}). A broad absorption band due to the Si-O stretching mode 
in amorphous silicates, centered at 9.7\,$\mu$m, is also commonly 
detected in galaxies.
Since the center of the silicate absorption coincides with a gap 
between the 6.2--8.6\,$\mu$m and 11.3--12.8\,$\mu$m PAH complexes, 
it is not readily apparent whether a 9.7\,$\mu$m flux minimum should
be interpreted as the ``trough'' between PAH emission features or as 
strong silicate absorption, or as a combination of the two.

In spectra observed towards heavily extincted Galactic lines of 
sight, like deeply embedded massive protostars, a strong silicate 
feature is often accompanied by mid-IR ice absorption features due
to molecules, frozen in grain mantles 
(Whittet et al. \cite{Whittet96}; Keane et al. \cite{Keane}). Among 
the simple ice molecules that have been identified in these grains 
are: H$_2$O, CO$_2$, CH$_3$OH, CO and CH$_4$ ice.
The first detections of interstellar ices in external galaxies 
were reported by Spoon et al. (\cite{Spoon00}) and Sturm et al. 
(\cite{Sturm}) in the nuclear spectra of NGC\,4945, M\,82 \& NGC\,253.
At the low resolution of the ISO-PHT-S spectrophotometer a deep and 
broad 3.0\,$\mu$m water ice feature, with the red wing extending 
far beyond the 3.3\,$\mu$m PAH feature, is clearly visible in the
spectrum of NGC\,4945. Also identified in the spectrum of NGC\,4945
are absorptions due to 4.26\,$\mu$m CO$_2$ ice and the unresolved 
blend of 4.62\,$\mu$m `XCN' with 4.67\,$\mu$m CO ice and 
4.6--4.8\,$\mu$m CO gas phase lines.

The first detections of 6--8\,$\mu$m ices in extragalactic sources
were made in the strongly absorbed mid-IR spectra of NGC\,4418 
(Spoon et al. \cite{Spoon01}) and IRAS\,00183--7111 (Tran et al. 
\cite{Tran}). Especially rich is the spectrum of NGC\,4418, which
displays absorption features due to 6.0\,$\mu$m water ice, 
6.85\,\&\,7.3\,$\mu$m hydrogenated amorphous carbons (HAC) and
7.67\,$\mu$m CH$_4$ ice, accompanied by a very deep 9.7\,$\mu$m 
silicate feature.

Encouraged by these findings we have searched our database
of $\sim$250 galaxies observed spectroscopically by ISO for galaxies
showing similar 6--8\,$\mu$m absorptions. In this paper we present
the outcome of this search, which resulted in a sample of 18 galaxies
showing evidence for the presence of 6.0\,$\mu$m water ice. Sect.\,2 
describes the sample. In Sect.\,3, we discuss the complex interplay 
of 6.0\,$\mu$m water ice and 5.25, 5.7, 6.25\,$\mu$m PAH emission
as well as the effect of redshift on the detectability of 
the blue wing of the water ice feature in ISO data. Sect.\,4 describes 
the classification of the ice galaxies. In Sect.\,5, we present the 
absorption and emission profile analysis. The results are discussed 
in Sect.\,6. Conclusions are stated in Sect.\,7.

\begin{figure*}[]
 \begin{center}
 \begin{picture}(400,350)(0,0)
  \psfig{figure=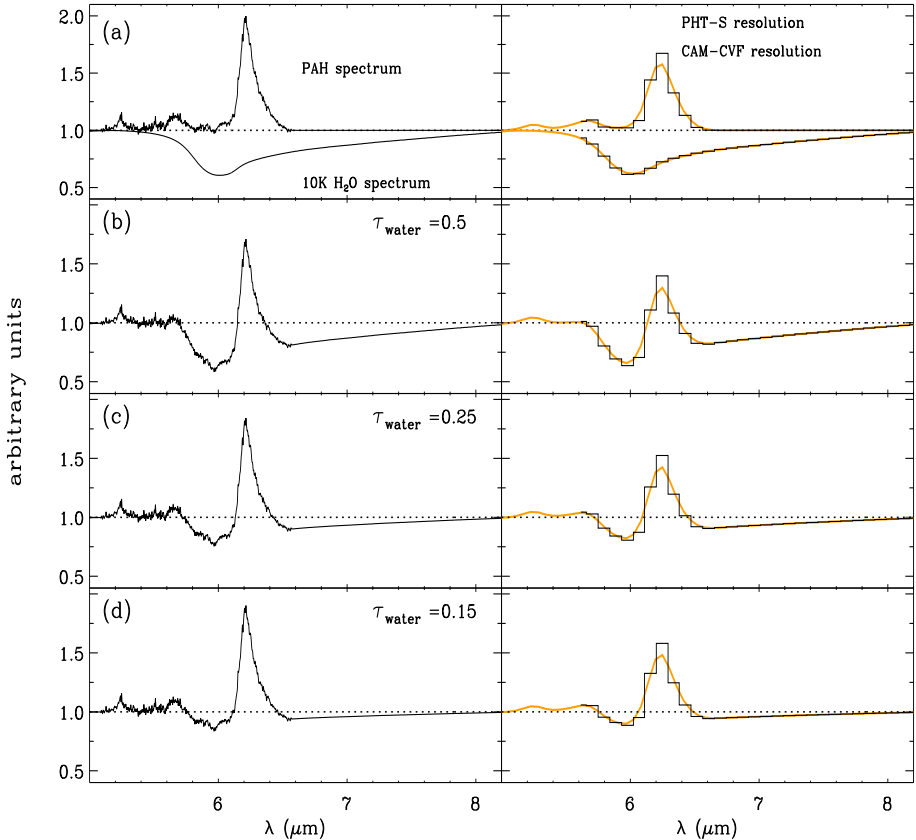,width=400pt,height=350pt,angle=90}
 \end{picture}
\end{center}
\caption{A sequence of plots illustrating the effect of rebinning a 
pseudo high resolution spectrum (left panels) to the lower resolution 
of ISO-PHT-S and ISO-CAM-CVF (right panels). A redshift of cz=12000 is 
adopted and the grey scale represents the ISO-CAM-CVF rebinned spectra. 
The pseudo spectrum is composed of 6.0 $\mu$m H$_{2}$O ice absorption 
and PAH emission bands found in the 5--7 $\mu$m range, superimposed on 
a flat continuum as shown in panel (a). In the remaining panels, 
(b, c, and d), the relative contributions of the H$_{2}$O absorption 
and PAH emission profiles are varied to investigate the behaviour of 
the spectral signatures. Note that we have discarded the 7.7\,$\mu$m 
PAH feature in our toy model. This strong feature would start 
contributing to the spectrum at around 7\,$\mu$m.}
\label{waterpah}
\end{figure*}

\begin{figure*}[]
 \begin{center}
  \psfig{figure=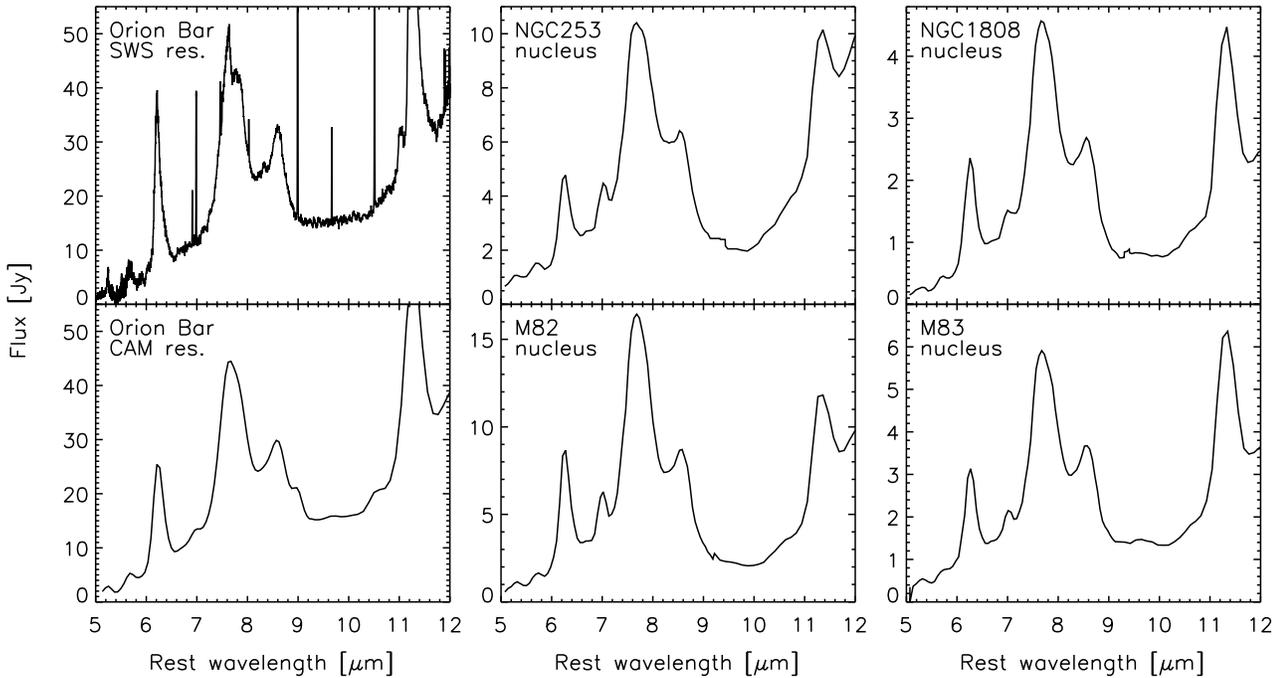,angle=0}
 \end{center}
\caption{A collection of spectra showing strong 5.25 and 5.7\,$\mu$m 
PAH features (Class 6 galaxies; see Sect.\,4). The four galaxies shown 
in the middle and right panels
were observed using ISO-CAM-CVF, at low spectral resolution. The left 
panels show the spectrum of the Orion Bar, as observed at high spectral 
resolution by ISO-SWS (upper left) and degraded to ISO-CAM-CVF 
resolution (lower left). Note that the four galaxies shown represent 
the strongest cases of 5.25\,$\mu$m and 5.7\,$\mu$m PAH emission. The 
average galaxy may have far lower 5.25\,$\mu$m/6.2\,$\mu$m and 
5.7\,$\mu$m/6.2\,$\mu$m PAH ratios.}
\label{5_6_um_pahs}
\end{figure*}

\section{Observations}

\subsection{ISO galaxy sample}

In order to study the mid-infrared spectral properties of galaxies
we have created a database of some 250 ISO galaxy spectra. The 
database has been compiled from observations performed with the
PHT-S, CAM-CVF and SWS-01 instrument modes of ISO. The database
comprises:
\begin{itemize}
\item Nearly all ISO-PHT-S galaxy observations from a variety of 
programs, except for galaxies at R.A. 13--24\,hrs from the 
``Normal Galaxy'' program (P.I. G.\,Helou).
\item Four ISO-SWS-01 spectra of nearby galactic nuclei (Sturm 
et al. \cite{Sturm}).
\item ISO-CAM-CVF spectra from the ZZULIRG proposal (Tran et al. 
\cite{Tran}), as well as the active galaxy spectra discussed by
Laurent et al. (\cite{Laurent}).
\end{itemize}

By nature, this sample is not complete in any strict statistical 
sense.

\subsection{Selected sample}

We have defined a subsample of 
galaxies with good S/N spectra and sufficient wavelength coverage
blueward of 6.0\,$\mu$m. The criteria we applied were 
(S/N)$_{6.5\mu {\rm m}}$\,$>$8 and, for ISO-PHT-S in addition, 
cz$>$3000\,km/s. The resulting sample of 103 galaxies misses many of 
the observations of fainter ULIRGs and nearby galaxies in the ISO 
archive. It is dominated by Seyferts, which constitute nearly half 
the sample.

Within this sample we found 18 galaxies showing 6.0\,$\mu$m 
H$_2$O ice absorption (Figs.\,\ref{class1}---\ref{class3b}),
including NGC\,4418 (Spoon et al. \cite{Spoon01}). 
We also looked for galaxies showing the opposite: no sign of 
water ice up to $\tau_{\rm ice}<$0.1--0.3 along the line 
of sight. Within our sample of 103 good S/N galaxy spectra, 28 
galaxies fulfil this criterion. Examples of 
ice-poor galaxies are shown in Figs.\,\ref{5_6_um_pahs}, 
\ref{sil_no_ice_panel}\,\&\,\ref{no_sil_no_ice_panel}.
For the remaining 57 out of 103 galaxies, the quality of the spectra
is either not good enough to detect the presence of a 6.0\,$\mu$m 
water ice feature (a higher S/N criterion would have biased the
sample against PAH-dominated spectra, which are weaker at 6--7\,$\mu$m
than continuum dominated AGN spectra), or to establish an upper limit 
better than 0.3 for the amount of water ice.

In the following Sections we show spectra for 32 galaxies. These spectra 
were obtained using the ISO-CAM-CVF (Cesarsky et al. \cite{Cesarsky}), 
ISO-PHT-S (Lemke et al. \cite{Lemke}) and ISO-SWS (De Graauw et al. 
\cite{DeGraauw}) spectro(photo)meters aboard ISO (Kessler et al. 
\cite{Kessler}).

The ISO-CAM-CVF spectra of galaxies I\,00183--7111, I\,00188--0856
and Arp\,220 have been taken from Tran et al. (\cite{Tran}). 
The ISO-CAM-CVF spectra of NGC\,253, NGC\,613, NGC\,1068, M\,82, 
NGC\,1365, NGC\,1808, NGC\,4945, M\,83, Circinus and I\,23128--5919 
have been taken from Laurent et al. (\cite{Laurent}). 

The ISO-PHT-S spectra of galaxies Mrk\,334, NGC\,23, I\,01173+1405, 
NGC\,828, I\,04385--0828, I\,05189--2524, MGC\,8--11--11, I\,06035--7102, 
UGC\,5101, Mrk\,231, Mrk\,273, Mrk\,279, I\,15250+3609, Arp\,220, 
I\,17208--0014, PKS\,2048--57, I\,20551--4250 and I\,23128--5919
have been reduced using standard routines of PIA\footnote{PIA is a joint 
development by the ESA Astrophysics Division and the ISO-PHT Consortium}
versions 8.1--9.0.1. 
The ISO-PHT-S spectra of NGC\,4418 and NGC\,4945 have been taken from 
Spoon et al. (\cite{Spoon01}) and Spoon et al. (\cite{Spoon00}), 
respectively.

The ISO-SWS spectrum of the nucleus of NGC\,1068 has been taken from 
Sturm et al. (\cite{Sturm}), that of the Orion Bar from Peeters et al. 
(\cite{Peeters}), and, finally, that of the nucleus of Circinus from
Moorwood et al. (\cite{Moorwood}).

\section{Ice absorption and PAH emission in the 5--7\,$\mu$m region}

Before discussing the presence and strength of ice features in the 
sources of our sample, we illustrate the interplay of emission and 
absorption using a simple toy model. This is necessary since the 
6\,$\mu$m ice feature is not always seen against a well-defined 
continuum. The `PAH' emission features at 7.7, 6.2, 5.7, and 
5.25\,$\mu$m may be present as well, leading to possible confusion
of minima between emissions with absorption. In addition, the wavelength 
coverage of the ISO data limits the accessible range, in particular for
low redshift objects observed with ISO-PHOT-SL (starting at 5.84\,$\mu$m).
Our model produces high resolution spectra
($R\,{\rm =\,} \lambda/\Delta \lambda \sim$\,1000) of the 5--7\,$\mu$m
spectral region, comprising an \h2o ice absorption feature and PAH 
emission bands superimposed on a continuum. Here, we describe the various 
components of the model, the consequences of rebinning to the lower 
\pht and \cam resolution ($R$\,$\sim$\,40--90) and also explore the 
effect that extinction may have on the resulting spectra. In addition, 
we produce model-spectra at different redshifts in order to investigate 
the effect that this has on the spectral signatures, in particular the 
relationship between redshift and wavelength coverage, which is 
particularly important for \pht spectra.

\subsection{Components of the model}

The components chosen for the model closely mimic absorption (ice) and 
emisson (PAH) features seen toward a variety of Galactic objects that 
have been extensively studied at high resolution. 
The spectra are modelled as:
F${\rm_{continuum} \times exp[-\tau(H_2O\,ice)]}$ + PAHs.
In order to allow easy comparison of the model generated pseudo-spectra, 
a false flat continuum is adopted. In Sect.\,5 the flat continuum is 
replaced by continuum choices fitting the individual observed spectra.
Our model ignores any contribution from the PAHs to the overall continuum.
This so-called PAH continuum can be seen in for instance the spectrum of 
the reflection nebula NGC\,7023 (Moutou et al. \cite{Moutou}) as the 
continuum extending below the PAH features. Since the PAH continuum 
only serves to dilute the other continuum and its associated ice feature, 
we ignore it here. 

Studies of Galactic star-forming regions, in 
particular high mass embedded protostars, with the short wavelength
spectrometer (SWS) of ISO show strong absorption features 
centred around 3.0\,$\mu$m and 6.0\,$\mu$m attributed to amorphous
\h2o ice. The spectral characteristics of \h2o ice (whether pure or in
various mixtures) have been well studied in the laboratory (Hagen et
al. \cite{Hagen}; Hudgins et al. \cite{Hudgins}; Maldoni et al. 
\cite{Maldoni}). The spectral changes that occur during warm-up from 
10\,K to 120\,K are irreversible. In the case of
the 6.0\,$\mu$m H$_2$O band, there is little spectral difference during
warm-up from 10\,K to 50\,K though the band begins to broaden
slightly and around 80\,K this effect is becomes discernible. 
This broadening reflects an annealing of the amorphous structure.
Further thermal cycling between 120\,K and 10\,K produces reversible 
effects due to thermal shinking and expansion of the ice lattice 
which causes more minor variations in the width of the bands,
particularly at 6.0\,\mum.
Beyond 120\,K the ice lattice transforms to polycrystalline ice.
This is particularly noticeable
in the 3.0\,$\mu$m H$_2$O band as sharp structure begins to appear. 
Though the bulk of the interstellar 3.0 and 6.0\,\mum
~absorption bands have been attributed to amorphous \h2o ice, there is
also evidence of additional molecules contributing to this feature in
a number of sources (Dartois \& d'Hendecourt 2001; Keane et
al. \cite{Keane}). As a conseqence of this, the astronomical
6.0\,$\mu$m feature varies somewhat from source to source and
therefore is not well suited to our simple model. Given the lack of 
3.0\,$\mu$m H$_2$O 
observational data for our sample, we have elected to represent the 
6.0\,$\mu$m ice absorption profile by a laboratory 
spectrum of pure amorphous \h2o ice at 12\,K. This profile has been
shown to satisfactorily reproduce the 6.0\,$\mu$m H$_2$O feature 
observed toward a number of Galactic young stellar objects (Keane et
al. \cite{Keane}). The left panel of  Fig.\,\ref{waterpah}a 
shows the profile of the 12\,K \h2o feature (database of the Sackler
Laboratory for Astrophysics in Leiden: {\tt
http://www.strw.leidenuniv.nl/$\sim$lab/}). The most striking aspect
of this profile is the broad long wavelength wing which   
extends at least as far as 8.0\,$\mu$m. There is little or no variation
in the \h2o profile for temperatures \ltsim 50\,K. For temperatures
greater than this the broad wing becomes deeper and extends well
beyond 8.0\,$\mu$m. The right panel of Fig.\,\ref{waterpah}a shows the
\h2o spectrum after it has been rebinned to the lower resolution of
\pht (histogram) and \cam (grey line), assuming a cz=12000\,km\,s$^{-1}$. 
The profile is of course very similar but for the \pht spectrum there 
is incomplete pixel coverage of the short wavelength wing when compared 
to the \cam spectrum.
The dominant PAH emission features (6.2, 7.7, 8.6, 11.3\,$\mu$m), 
observed toward a variety of Galactic sources, show
a wide degree of variation in the detailed profile shapes and relative
intensities (Peeters et al. \cite{Peeters}). On the other hand,
observations of PAH emission bands in external galaxies reveal very little
variation in the profiles (Rigopoulou et al. \cite{Rigopoulou};
Helou et al. \cite{Helou}). Since one of the main aims of the model 
is to investigate the effect of rebinning spectral features to the 
lower resolutions of \pht and ISO-CAM-CVF, adopting extra-galactic PAH 
emission profiles is inappropriate. Instead, high resolution Galactic PAH 
emission band profiles are adopted. The ISO-SWS PAH emission
spectrum of the Orion bar PAH is chosen for the model as this 
is a good representation of a region of active star-formation (Peeters
et al. \cite{Peeters}). The left panel of Fig.\,\ref{waterpah}a shows the
high resolution spectrum of the Orion Bar. Apart from the typical
6.2\,$\mu$m PAH CC stretching mode, the observed spectrum also shows
evidence for two weak PAH emission features near 5.25\,$\mu$m and
5.7\,$\mu$m (left panel of Fig.\,\ref{waterpah}a). The nature of these
weak PAH features is not very well studied but they are believed to be
combinations or overtone bands involving C-H bending vibrations
(Allamandola et al. \cite{Allamandola}). The strength of these weak
bands relative to the strong 6.2\,$\mu$m band is variable. 
Of a sample of 35 compact HII regions studied by Peeters et al. 
(in prep.) roughly 10\% of the sources show evidence for these weak 
PAH bands. Consequently, because of their variable strength, their 
presence is 
not always assured in spectra showing the dominant PAH bands and
there is no rule of thumb defining an expected ratio for the intensity
of these weak PAH bands to the 6.2\,$\mu$m band. 
The Orion bar is among the Galactic sources with the strongest 5.25 
and 5.7\,$\mu$m features relative to the 6.2\,$\mu$m feature. 
Comparison with high S/N CAM-CVF spectra of the 
brightest nearby starbursts suggests it to be fairly representative 
for those objects however (Fig.\,\ref{5_6_um_pahs}, confirmed by SWS 
spectroscopy of M\,82 and NGC\,253 (Sturm et al. \cite{Sturm})). 
We hence adopt the Orion Bar as a template for the 5.25, 5.7, and 
6.2\,$\mu$m PAH features. The PAH spectrum in Fig.\,\ref{waterpah} 
was obtained by subtracting a spline continuum from its spectrum. 
Fig.\,\ref{waterpah} then suggests that in the presence of strong PAH 
emission, a shallow 6.0\,$\mu$m ice absorption of less than about 
10\% of the peak height of the 6.2\,$\mu$m PAH feature is very 
difficult to discriminate from the minimum between the 5.7 and 
6.2\,$\mu$m PAH features, especially for the limited wavelength coverage 
of ISO-PHT-S. Finally, the 7.7 and 8.6\,$\mu$m PAH bands have been 
clipped from the spectrum as they will be dealt with separately 
(Spoon et al. in prep.).

The right panel of Fig.\,\ref{waterpah}a shows the Orion-Bar PAH
features rebinned to the lower resolution of \pht (histogram) and \cam
(grey line). In rebinning, the peak height of the 6.2\,$\mu$m feature 
is reduced, more for ISO-CAM-CVF than for ISO-PHT-S. In the \cam 
spectrum, the weak PAH bands are still present, though they are
flattened and broaded as compared to the high resolution
spectrum. Only the peak of the 5.7\,$\mu$m weak PAH feature is present
in the \pht spectrum since at a cz of 12000\,km\,s$^{-1}$ the
first pixel is at 5.6\,$\mu$m. This is discussed further in Sect.\,3.3. 
Finally, the location of the water (i.e., in front,
behind or mixed with the PAHs) had little effect on the resulting
shape of the spectral features, in particular on the profile of the 
6.2\,$\mu$m feature. We hence ignored extinction on the PAHs in our 
toy model. Detailed fits to the extragalactic spectra are presented 
in Sect.\,5, in which the flat continuum is replaced by continua 
determined individually for each galaxy.

\subsection{Effects on the model of varying the component contributions 
and extinction}

To investigate the interplay of the \h2o absorption feature and the 
6.2\,$\mu$m PAH emission bands, the relative contribution of each 
component is varied. This is shown in Fig.\,\ref{waterpah}b--d, where
the left panel is the high resolution input spectrum and the right
panel represents the rebinned low resolution pseudo spectra of the
model. The dotted line indicates the flat continuum. Panel (b)
illustrates the effect of a strong \h2o band on the PAH emission
bands. Regardless of the fact that the 6.2\,$\mu$m features lies within 
the long wavelength wing of the \h2o feature, its profile is preserved
(left panel). The 5.7\,$\mu$m weak PAH band, on the other hand, is lost 
within the short wavelength wing, though there is a slight hint of a 
peak at 5.7\,$\mu$m. The 5.25\,$\mu$m PAH feature is still present as
the \h2o profile does not reach as far as this short wavelength position. 
Rebinning to the low resolution of \pht and \cam (right panel) shows that 
the \h2o feature, in combination with the 6.2\,$\mu$m PAH emission
band, dominate the model spectrum and there is little if no evidence
for the weak PAH bands in the \pht spectrum. Only the \cam spectrum
reveals the presence of the 5.25\,$\mu$m weak PAH band, due to the
wider wavelength coverage. As the strength of the \h2o absorption
band is reduced (Panels (c) and (d)), the 6.2\,$\mu$m feature starts to
dominate the model spectra and now only weakly sits within the long 
wavelength wing of the \h2o feature. Also, the spectral signature of
the 5.7\,$\mu$m weak PAH band starts to become pronounced and in Panel 
(d) the structure is clearly distinguished lying above the flat
continuum. The crucial result of this study is that the profile of the
6.2\,$\mu$m PAH band is little affected by the presence of an \h2o
ice absorption feature. In addition, it is evident from this analysis that 
the weak PAH bands do not influence the overall characteristics of
the structure seen in the model spectra. However, the weak PAH
emission features may enhance/mimick the effects of weak ice absorption.

\begin{figure}[]
 \begin{center}
   \psfig{figure=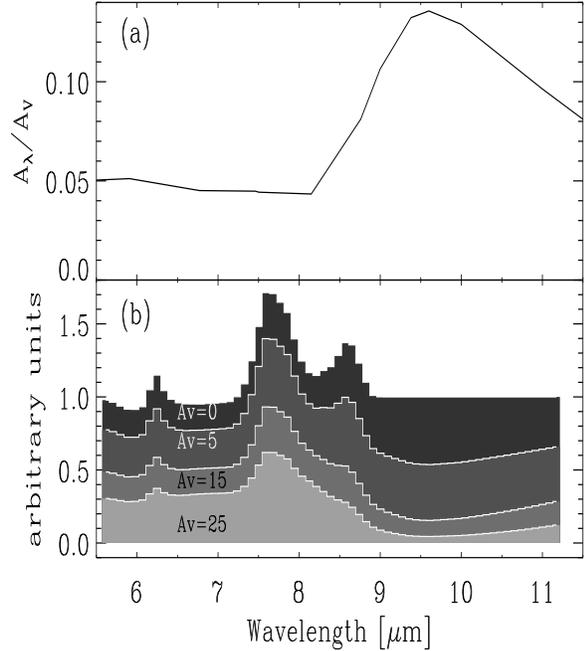,width=8.8cm,height=10cm,angle=90}
 \end{center}
\caption{Panel (a) shows a mid-IR extinction law of the Galactic centre
derived from hydrogen recombinatin lines (Lutz \cite{Lutz99}). Illustrated 
in panel (b) is the effect of extinction on the model spectra for
different amounts of {\it A}$_{\rm v}$, assuming the above extinction
curve. The effect is most severe for the 8.6\,$\mu$m band as all
profile information becomes lost within the 10\,$\mu$m silicate feature.}
\label{extinctionlaws}
\end{figure}

The possible effect that extinction may have on the spectral features
of the pseudo spectra were also investigated. In light of the fact
that the Galactic Center has been well studied and also because of the
homogeneous nature of the foreground extinction, the extinction law
derived by Lutz (\cite{Lutz99}) for the Galactic Centre is adopted
(Fig.\,\ref{extinctionlaws}a). We have also investigated the effect of
other extinction laws (Tran et al. \cite{Tran}; Draine \cite{Draine}) 
but they are substantially the same and
only the effects of the Lutz Law are discussed in detail here.
The extinction is applied as if the
continuum+PAH+ice combination were behind a column of material. Though
the situation is less clear for external galaxies, this approach
allows for a qualitative assessment of the effect of extinction on the
model spectra. In order to clearly demonstrate the effect of
extinction, the 7.7 and 8.6\,$\mu$m PAH bands of the Orion Bar have
been included. The top plot in Fig.\,\ref{extinctionlaws}b shows the
unextincted ({\it A$_v$}=0) model spectrum. As the degree of
extinction increases from {\it A$_v$}=0 to {\it A$_v$}=25 the
8.6\,$\mu$m PAH band is "eaten away" by the short wavelength wing of the
10\,$\mu$m silicate feature. Eventually, the depth of the silicate band
is so great that the 8.6\,$\mu$m feature is completely lost within this
band. The same effect, but with less severity, is noticeable for the
7.7\,$\mu$m emission feature. In the case of the 6.2\,$\mu$m band, the
peak intensity decreases as the degree of extinction increases, but
information on the profile shape remains intact and distinguishable. 
Thus, the spectral characteristics of features shortward of 7\,$\mu$m
are unaffected by the effects of extinction. Furthermore, due to the 
sensitivity of the features beyond 7\,$\mu$m to the degree of
extinction they have been neglected in the model.

\begin{figure}[]
 \begin{center}
 \begin{picture}(250,250)(0,0)
  \psfig{figure=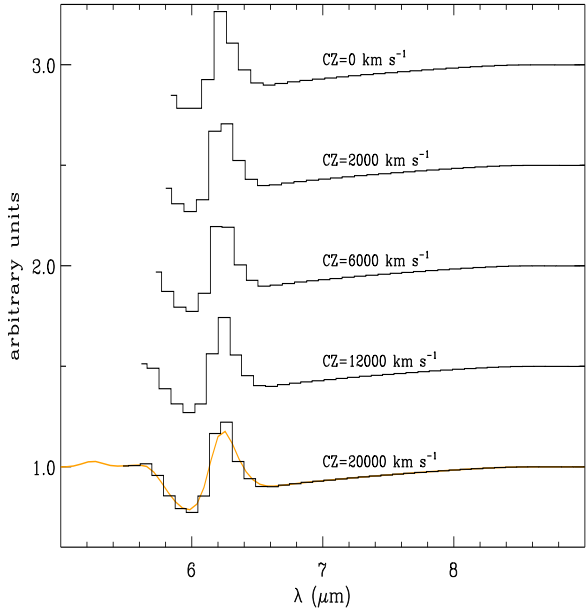,width=250pt,height=250pt,angle=90}
 \end{picture}
 \end{center}
\caption{A comparision of the spectral appearance of panel (b) from 
Fig.\,\ref{waterpah} rebinned to ISO-PHT-S resolution as a function of 
redshift. At low redshifts the presence of 6.0 $\mu$m H$_{2}$O ice absorption 
is not very convincing. However, as the redshift is increased the starting 
position of the first pixel shifts to lower wavelengths resulting in more of 
the 6.0 $\mu$m H$_{2}$O profile being clearly discernible. Since the wavelength
coverage of ISO-CAM-CVF (grey line) is significantly larger than that of 
ISO-PHT-S, there is not much variation in wavelength of the first pixel with 
redshift.}
\label{redshift}
\end{figure}

\subsection{Profile appearance as a function of redshift}

We have also investigated the effects of redshift combined with the 
limited spectral coverage on the appearance of the spectrum.
A redshift (cz) range of 0\,km\,s$^{-1}$ to 20\,000\,km\,s$^{-1}$ was 
chosen as this is representative of the cz range of the observed 
galaxies (Table\,\ref{optdepths}). 
Fig.\,\ref{redshift} illustrates the spectral variation as a
function of redshift. The high resolution spectrum shown in panel (b) of 
Fig.\,\ref{waterpah} was adopted as the input to the rebinning routines and as 
before the histogram and the grey line represent the \pht and \cam 
rebinned data, respectively. Since the spectral coverage of \cam extends
down as far as 5\,$\mu$m there is not much variation in the position of the 
first pixels as a function of redshift. Hence in Fig.\,\ref{redshift} only 
one \cam spectrum is shown. For redshifts of 6000\,km\,s$^{-1}$ or 
greater, the 6.0\,$\mu$m \h2o absorption profile is readily seen in the model
\pht ~spectra. On the other-hand, the presence of \h2o
is not so obvious in \pht ~spectra for redshifts less than
6000\,km\,s$^{-1}$. As the redshift decreases the pixel coverage of
the short wavelength wing of the \h2o absorption feature diminishes
and only one or two pixels begin to rise at the shortest wavelengths. 
Consequently, there is not much real evidence for the blue wing of the 
\h2o band even though the original high resolution model spectrum
clearly shows strong \h2o ice absorption (panel (b) in
Fig.\,\ref{waterpah}). In the case of the 6.2\,$\mu$m PAH emission
band, the only notably change is that the rebinned structure of the
profile peak varies with redshift.

\section{Classification}

\begin{figure*}[]
 \begin{center}
  \psfig{figure=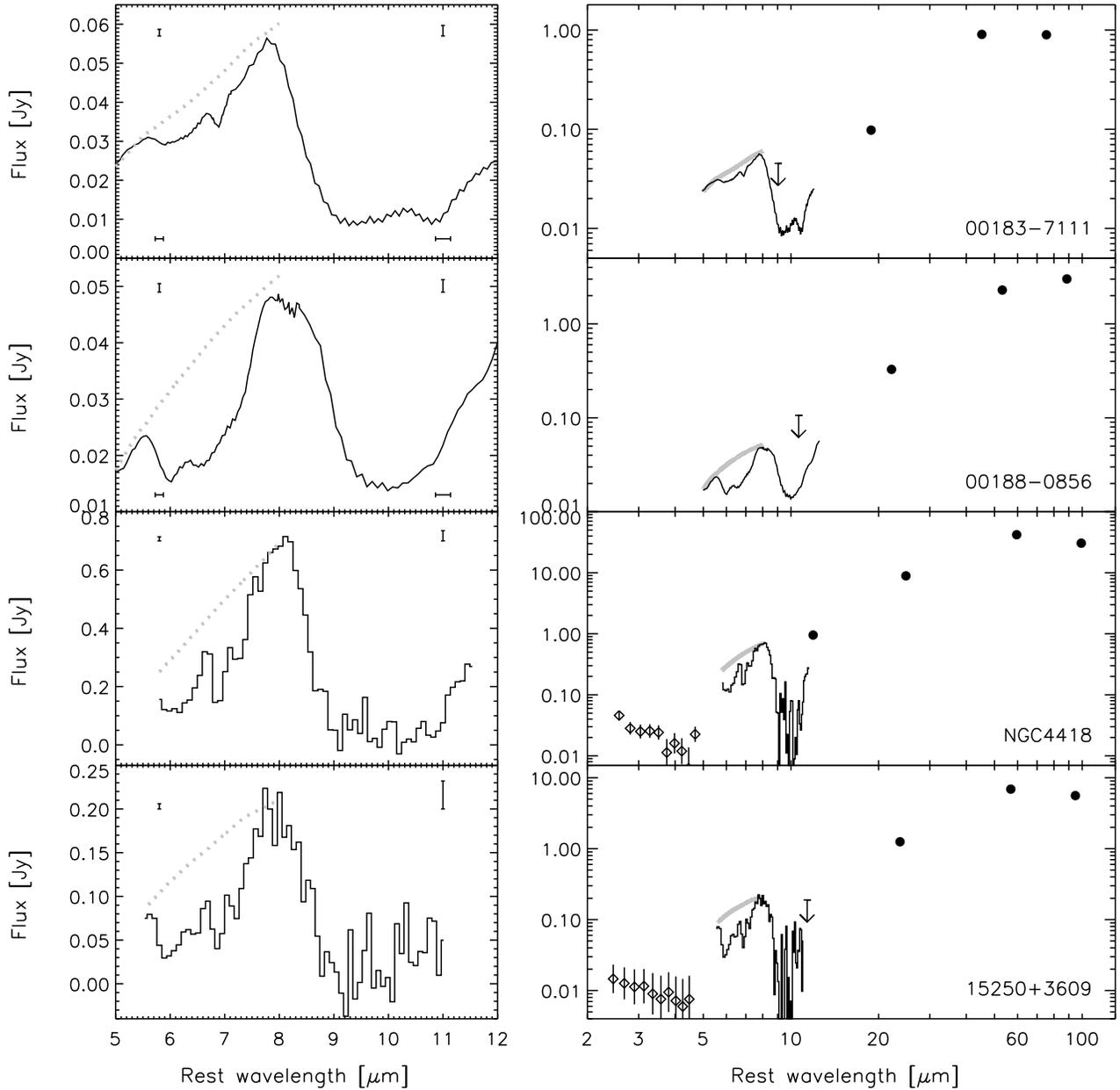,angle=0}
 \end{center}

\vspace*{-7mm}

\caption{Class-1 mid-IR galaxy spectra. Galaxies in Class 1 exhibit
a clear 6.0\,$\mu$m ice absorption feature, without clear evidence for
the presence of a 6.25\,$\mu$m PAH feature. All sources in this class
have in common that they have a broad continuum flux peak near 
7.7\,$\mu$m, clearly broader than a 7.7\,$\mu$m PAH feature. The panels 
on the left show the 5--12\,$\mu$m spectra on a linear scale, while the
panels on the right show the same spectra on logarithmic scale, with
smoothed and rebinned 2--5$\mu$m ISO-PHT-S spectra and 12-100\,$\mu$m 
IRAS fluxes added. All fluxes have been k-corrected. The thick grey lines 
show what the 5.5--8.0\,$\mu$m continuum would be like if no ice absorption
were present. The vertical error bars at 5.8 and 11\,$\mu$m denote the 
pixel to pixel flux error (1$\sigma$, so not $\pm$1$\sigma$), while the 
horizontal error bars denote the gradually changing ISO-CAM-CVF spectral 
resolution (R$\sim$40).}
\label{class1}
\end{figure*}

\begin{figure*}[]
 \begin{center}
  \psfig{figure=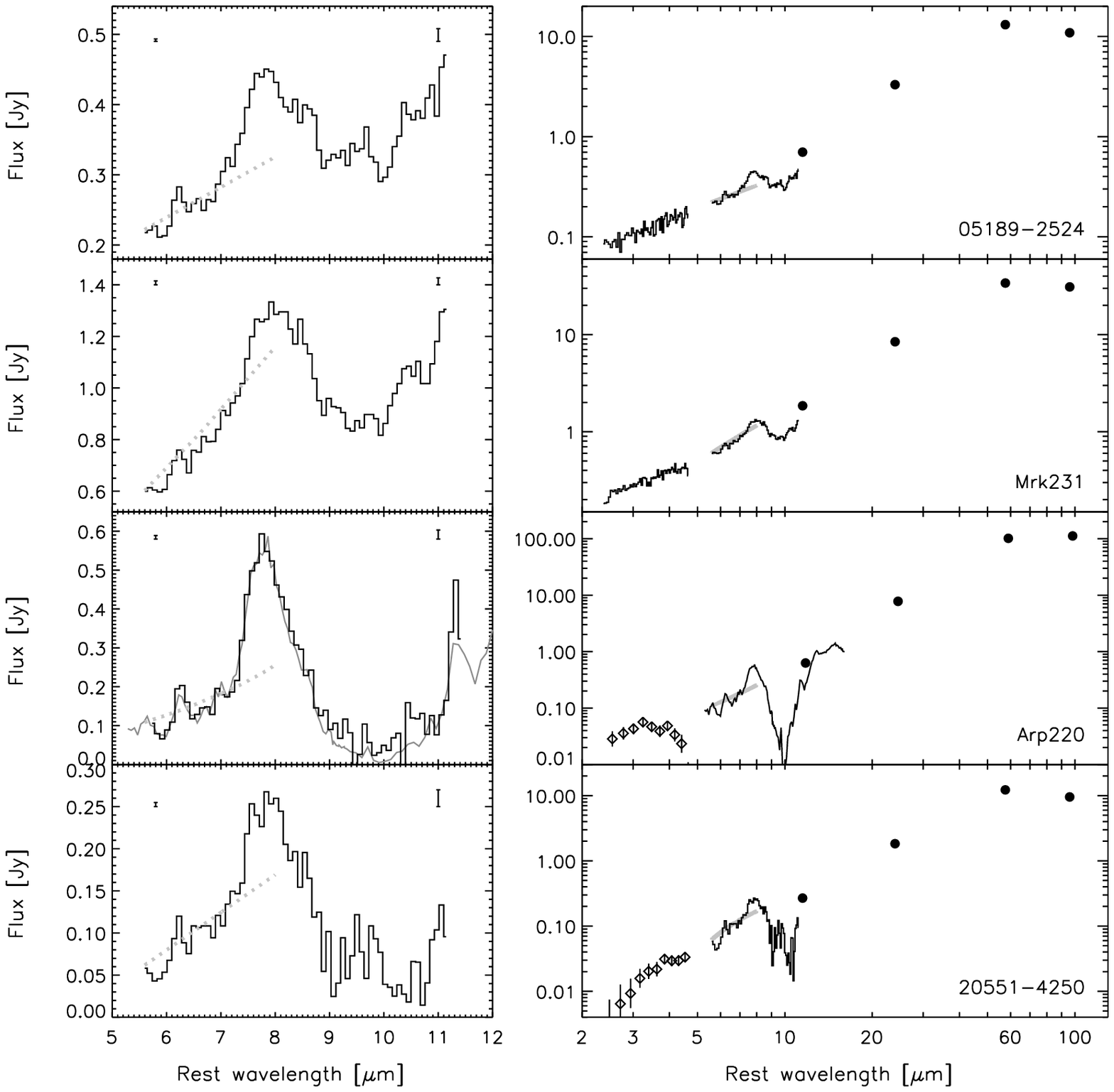,angle=0}
 \end{center}

\vspace*{-7mm}

\caption{Class-2 mid-IR galaxy spectra. Galaxies in Class 2 exhibit
a clear 6.0\,$\mu$m ice absorption feature, with clear evidence for
the presence of a 6.25\,$\mu$m PAH feature. All sources in this class
have in common that they have a broad continuum flux peak near 
7.7\,$\mu$m, clearly broader than a 7.7\,$\mu$m PAH feature. The panels 
on the left show the 5--12\,$\mu$m spectra on a linear scale, while the
panels on the right show the same spectra on logarithmic scale, with
smoothed and rebinned 2--5$\mu$m ISO-PHT-S spectra and 12-100\,$\mu$m 
IRAS fluxes added. All fluxes have been k-corrected. The thick grey lines 
show what the 5.5--8.0\,$\mu$m continuum would be like if no ice absorption
were present. For Arp\,220 both the ISO-PHT-S ({\it black}) and 
ISO-CAM-CVF ({\it grey}) spectrum are shown in the left panel.
The vertical error bars at 5.8 and 11\,$\mu$m denote the 
pixel to pixel flux error (1$\sigma$, so not $\pm$1$\sigma$).}
\label{class2}
\end{figure*}

\begin{figure*}[]
 \begin{center}
  \psfig{figure=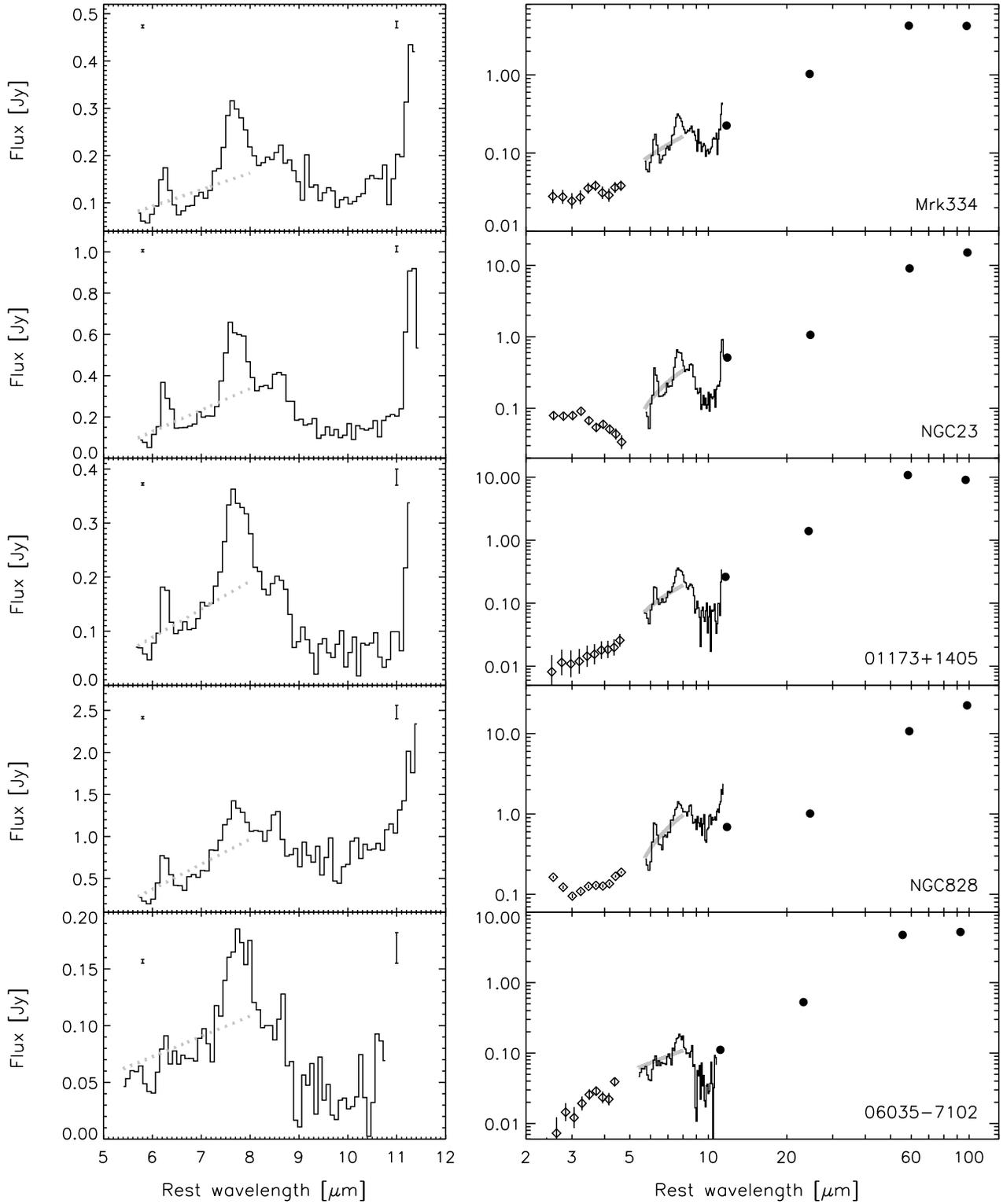,angle=0}
 \end{center}

\vspace*{-7mm}

\caption{Class-3 mid-IR galaxy spectra. Galaxies in Class 3 exhibit
a narrow 6.0\,$\mu$m ice absorption feature, partially filled in by
the wings of the 6.25\,$\mu$m PAH feature. All sources in this class
have in common that the flux peak near 7.7\,$\mu$m has the typical 
PAH width. 
The panels on the left show the 5--12\,$\mu$m spectra on a linear 
scale, while the panels on the right show the same spectra on 
logarithmic scale, with smoothed and rebinned 2--5$\mu$m ISO-PHT-S 
spectra and 12-100\,$\mu$m IRAS fluxes added. All fluxes have been 
k-corrected. The thick grey lines show what the 5.5--8.0\,$\mu$m 
continuum would be like if no ice absorption were present.
For NGC\,4945 both the ISO-PHT-S ({\it black}) and 
ISO-CAM-CVF ({\it grey}) spectrum are shown in the left panel.
The vertical error bars at 5.8 and 11\,$\mu$m denote the 
pixel to pixel flux error (1$\sigma$, so not $\pm$1$\sigma$).}
\label{class3a}
\end{figure*}

\begin{figure*}[]
 \begin{center}
  \psfig{figure=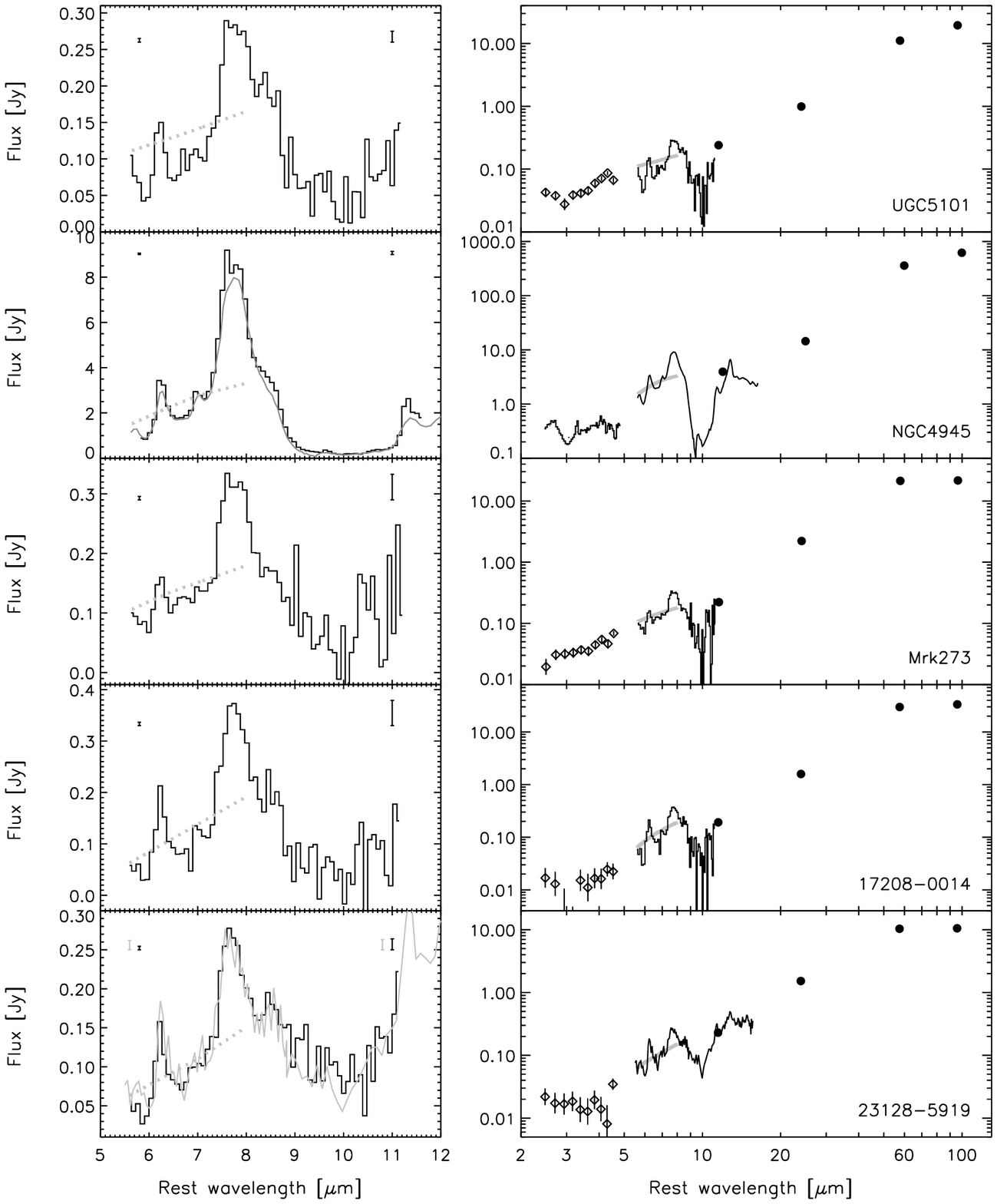,angle=0}
 \end{center}
\caption{Class 3 (continued)}
\label{class3b}
\end{figure*}

Although likely the spectra of galaxies represent a continuous
distribution with variable absorption and emission components,
we have classified the galaxies in our sample into six categories, 
based on their mid-IR spectral properties. The first three catagories
comprise the galaxies with evidence for the presence of 6.0\,$\mu$m 
water ice absorption in their spectra:

\begin{itemize}
\item {\bf Class 1}: These galaxies (Fig.\,\ref{class1}) exhibit a 
6.0\,$\mu$m H$_2$O ice absorption feature, without contamination by 
6.2\,$\mu$m PAH emission. Galaxies of this type also show other 
6--8\,$\mu$m absorption features -- most commonly the 6.85\,$\mu$m 
feature, attributed to HAC. Note that all galaxies in this class have 
a broad flux peak at $\sim$8\,$\mu$m. Judging from their complete
2--200\,$\mu$m SEDs, all Class 1 galaxies have a deep 9.7\,$\mu$m 
silicate feature. For the two galaxies for which 2--5\,$\mu$m 
ISO-PHT-S data is available, NGC\,4418 and IRAS\,15250+3609, there 
is a sharp increase in flux between 5 and 6\,$\mu$m. 
\item {\bf Class 2}: These galaxies (Fig.\,\ref{class2}) show
6.0\,$\mu$m H$_2$O ice absorption features, partially filled-in by
weak 6.2\,$\mu$m PAH emission. Other 6--8\,$\mu$m absorption 
features are not always as apparent. Note that like in Class 1, all 
galaxies in this class have a broad 8\,$\mu$m feature.
\item {\bf Class 3}: These galaxies (Fig.\,\ref{class3a}\&\,\ref{class3b})
exhibit a 6.0\,$\mu$m H$_2$O ice absorption feature, largely filled 
in by strong 6.2\,$\mu$m PAH emission. No other 6--8\,$\mu$m absorption 
features have been found. All galaxies in this class
show a 7.7\,$\mu$m PAH emission peak, clearly narrower than the 
broad 8.0\,$\mu$m feature seen in Class 1\,\&\,2 galaxies.
\end{itemize}

Table\,\ref{optdepths} lists the classification for all 18 ice galaxies.
For completeness we also list galaxies with 3.0\,$\mu$m ice absorption
in their nuclear spectra: 
NGC\,253, M\,82 and NGC\,4945 (Sturm et al. \cite{Sturm}; Spoon et al. 
\cite{Spoon00}). Only for NGC\,4945 has the usually far weaker 6.0\,$\mu$m 
feature also been detected -- at similar strength as the 3.0\,$\mu$m
one. This apparent inconsistency may be due to aperture size effects
(the 3.0\,$\mu$m feature was measured by ISO-PHT-S; the 6.0\,$\mu$m 
feature by ISO-CAM-CVF) or due to dilution by less obscured emission, 
filling in the water ice features.

\begin{figure*}[]
 \begin{center}
  \psfig{figure=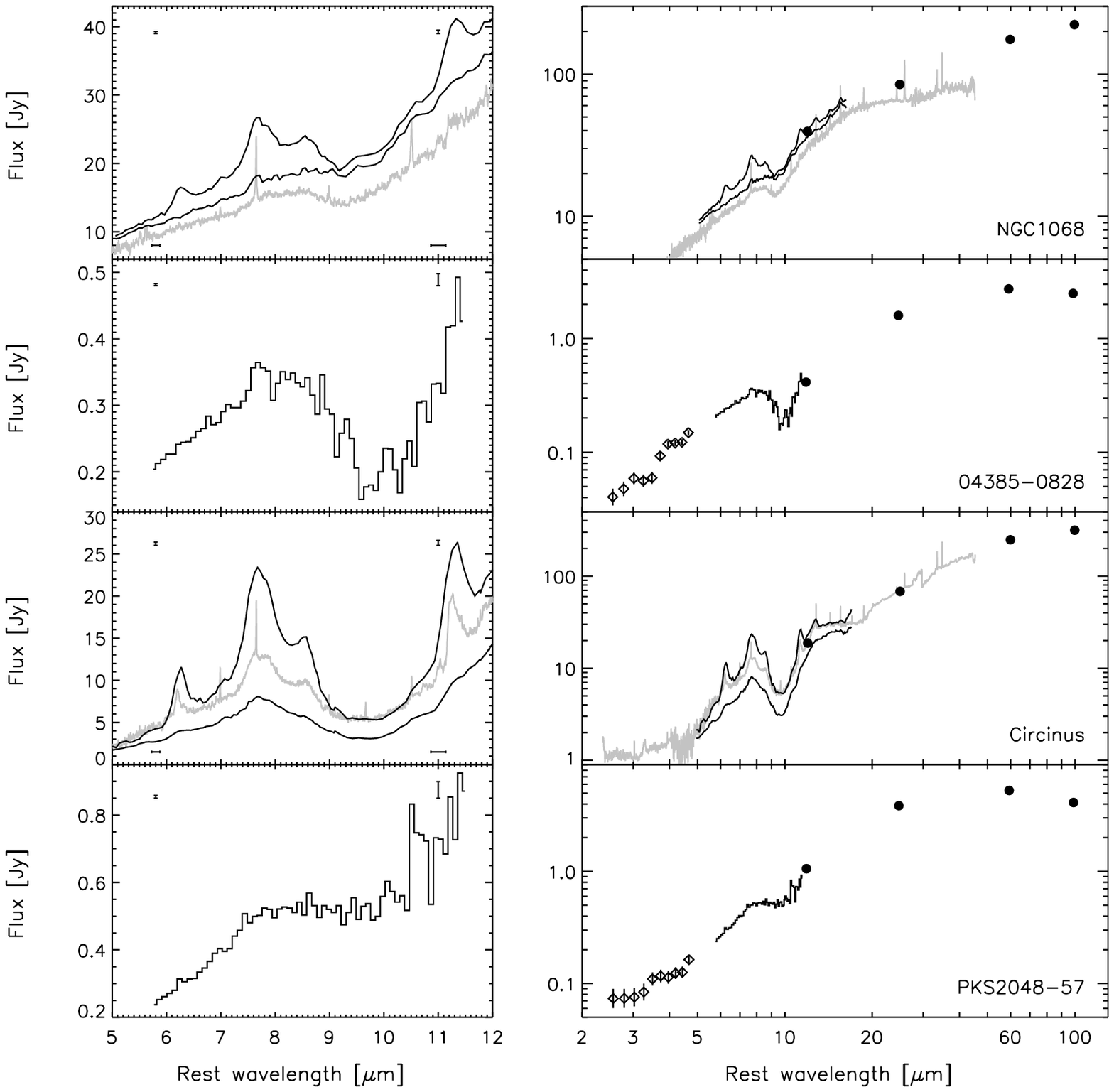,angle=0}
 \end{center}

\vspace*{-7mm}

\caption{Class 4 mid-IR galaxy spectra. Galaxies of this kind are 
ice-poor and show the presence of silicates along the line of sight. 
The presence of silicates is inferred from the absorption feature at 
9.7\,$\mu$m, characteristic for silicates. The upper limit on H$_2$O 
ice is inferred from the absence of a 6.0\,$\mu$m absorption feature. 
For NGC\,1068 
and for Circinus three spectra are shown: the ISO-SWS nuclear spectrum
in grey and the ISO-CAM-CVF nuclear and total spectra in black. 
The offset between the three spectra is most likely an aperture effect. 
The panels 
on the left show the 5--12\,$\mu$m spectra on a linear scale, while the
panels on the right show the same spectra on logarithmic scale, with
smoothed and rebinned 2--5$\mu$m ISO-PHT-S spectra and 12-100\,$\mu$m 
IRAS fluxes added. All fluxes have been k-corrected. The vertical error 
bars at 5.8 and 11\,$\mu$m denote the pixel to pixel flux error 
(1$\sigma$, so not $\pm$1$\sigma$), while the horizontal error bars 
denote the gradually changing ISO-CAM-CVF spectral resolution (R$\sim$40).}
\label{sil_no_ice_panel}
\end{figure*}

\begin{figure*}[]
 \begin{center}
  \psfig{figure=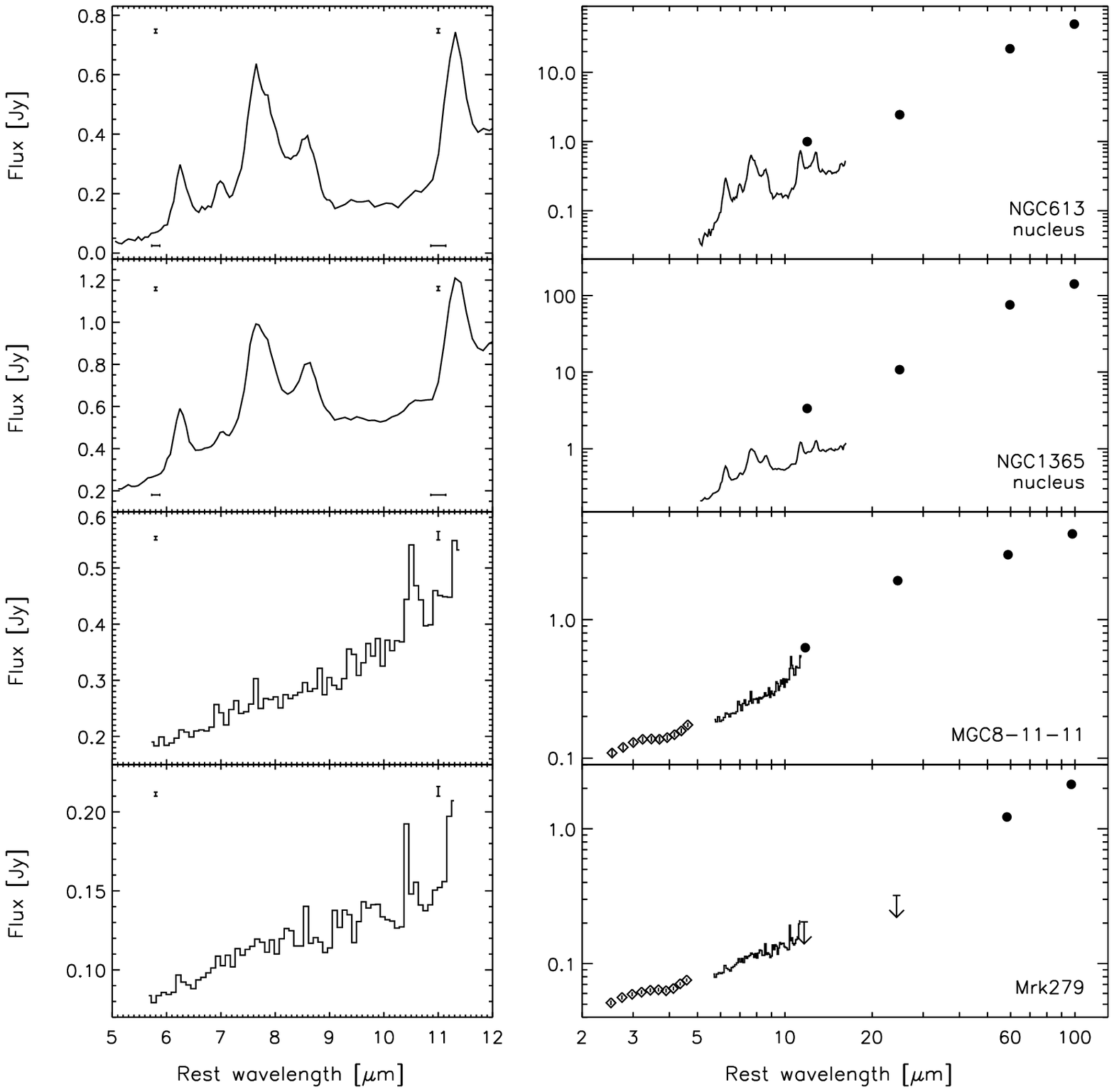,angle=0}
 \end{center}

\vspace*{-7mm}

\caption{Class 5 mid-IR galaxy spectra. Galaxies of this kind are 
ice-poor and show no clear evidence for the presence of silicates 
along the line of sight. Note the clear mismatch of the IRAS FSC fluxes 
with the ISO-CAM-CVF nuclear spectra for both NGC\,613 and NGC\,1365.
This most probably is an aperture-size effect. The presence of strong PAH
emission both short- and longward of 9.7\,$\mu$m makes it difficult to
completely rule out silicate absorption in the nuclei of NGC\,613 and 
NGC\,1365. Compared to Circinus and NGC\,520 however 
(Fig.\,\ref{sil_no_ice_panel}), the silicate depth must at least be small.
The panels on the left show the 5--12\,$\mu$m spectra on a linear scale, 
while the panels on the right show the same spectra on logarithmic scale, 
with smoothed and rebinned 2--5$\mu$m ISO-PHT-S spectra and 12-100\,$\mu$m 
IRAS fluxes added. All fluxes have been k-corrected. The vertical error 
bars at 5.8 and 11\,$\mu$m denote the pixel to pixel flux error 
(1$\sigma$, so not $\pm$1$\sigma$), while the horizontal error bars denote 
the gradually changing ISO-CAM-CVF spectral resolution (R$\sim$40).}
\label{no_sil_no_ice_panel}
\end{figure*}

Within our sample of 103 galaxies with good S/N spectra and sufficient
wavelength coverage we also recognize three other classes of spectra:

\begin{itemize}
\item{\bf Class 4:} These galaxies (Fig.\,\ref{sil_no_ice_panel}) show a
smooth featureless continuum in the 5--6\,$\mu$m range as well as 
definite signs of 9.7\,$\mu$m silicate absorption. The presence of
silicate absorption is most easily determined by interpolating
the 6\,$\mu$m ISO and the 12\,$\mu$m IRAS flux points (or in case
of ISO-CAM-CVF, using the 13--15\,$\mu$m continuum instead).
\item{\bf Class 5:} These galaxies (Fig.\,\ref{no_sil_no_ice_panel})
show a smooth featureless continuum in the 5--6\,$\mu$m range without
clear signs of 9.7\,$\mu$m silicate absorption. The absence of a clear
silicate absorption feature is most easily determined by interpolating 
the 6\,$\mu$m ISO and the 12\,$\mu$m IRAS flux points (or in case of 
ISO-CAM-CVF, using the 13--15\,$\mu$m continuum instead).
\item{\bf Class 6:} These galaxies (Fig.\,\ref{5_6_um_pahs}) clearly show
5.25 and 5.7\,$\mu$m PAH features, without clear signs of 9.7\,$\mu$m 
silicate absorption. The absence of a clear silicate absorption feature 
is most easily determined by interpolating the 6\,$\mu$m ISO and the 
12\,$\mu$m IRAS flux points (or in case of ISO-CAM-CVF, using the 
13--15\,$\mu$m continuum instead).
\end{itemize}

Table\,\ref{iceless_stats} lists the observed mid-IR features for the
13 out of 28 Class 4--6 galaxies displayed in this paper. It is hard to 
quantify consistently the limits on ice absorption in these three classes 
because of the presence of PAH features. For Class 4 and 5 we estimate 
$\tau_{\rm ice}<$0.1--0.3. For Class 6, in all spectra the presence of 
weak 5.7\,$\mu$m and strong 6.2\,$\mu$m features 
may easily mask ice features as strong as $\tau_{\rm ice}$=0.3.
We note that among the Class 6 galaxies, both NGC\,253 and M\,82 do 
have a weak 3.0\,$\mu$m water ice feature (Sturm et al. \cite{Sturm}),
but the 6.0\,$\mu$m ice limit is $\tau_{\rm ice}$=0.3 for both galaxies.

The total number of ISO galaxies classified into Classes 1--6 is small:
just 46 out of $\sim$250 galaxies in our sample. 
The combination of both a good coverage of the 5.5--6.0\,$\mu$m
range and a good S/N over 5.5--6.5\,$\mu$m proves to be a hard 
requirement to meet. With its superior sensitivity and better coverage 
of the 6\,$\mu$m region, SIRTF can be expected to refine our 
classification, find new members and provide a larger sample for 
statistical analyses.

\section{Absorption and emission profile analysis}

\subsection{Model fits}

The model described in Sect.\,3 is now applied to the Class 1--3 
sources presented in Sect.\,4. The flat continuum of Sect.\,3 is 
replaced by the 5--8\,$\mu$m continua derived individually for each 
source (thick grey lines in Figs.\,\ref{class1}\,--\,\ref{class3b}). For 
Class 2 and 3 sources, the continua were determined by interpolating 
the ISO-PHT-S short wavelength spectrum (2--5\,$\mu$m) linearly or 
logaritmically (depending on the SED shape) to the long wavelength 
IRAS data, ignoring the presence of PAH emission features. 
For Class 1 sources we followed another recipe. In the absence of
contamination by PAH emission bands, we assumed the 8\,$\mu$m flux 
peak to represent the local 8\,$\mu$m continuum (see Sect.\,6 for a 
discussion on the nature of the 8\,$\mu$m flux peak). For the three 
sources observed with ISO-CAM-CVF we then derived the 5--8\,$\mu$m
continuum by interpolating to the observed 5\,$\mu$m flux. For
NGC\,4418 and I\,15250+3609 we interpolated to the reddest 
2--5\,$\mu$m ISO-PHT-S data point, assuming a continuum shape 
similar to that of I\,00188--0856. Note that for all Class 1 sources
we included a 6.85\,$\mu$m absorption feature due to hydrogenated 
amorphous carbon (HAC; Furton et al. \cite{Furton}). A (strong) 
6.85\,$\mu$m absorption feature is commonly observed towards Galactic 
lines of sight (Chiar et al. \cite{Chiar}; Keane et al. \cite{Keane}). 

\begin{figure*}[]
 \begin{center}
  \psfig{figure=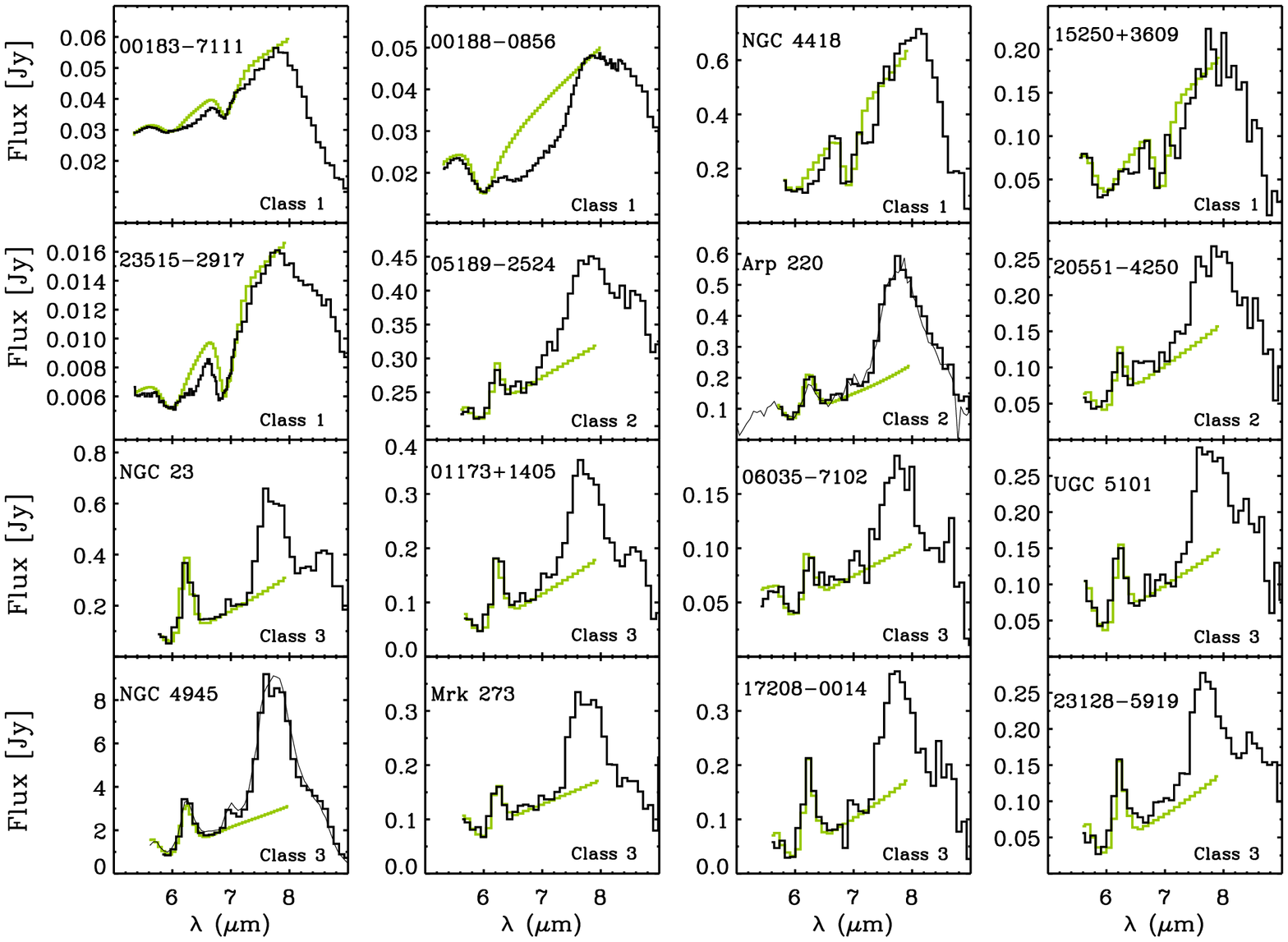,angle=90,width=19.cm,height=16cm}
 \end{center}
\caption{A comparison of all the sources with the best fitting models which 
contain both H$_2$O ice absorption and PAH emission bands. The grey 
histogram data represent the model fits. The three sources plotted in 
Fig.\,\ref{unsure_ice} are not shown here.}
\label{ice_fits}
\end{figure*}

\begin{figure*}[]
 \begin{center}
 \begin{picture}(400,300)(0,0)
  \psfig{figure=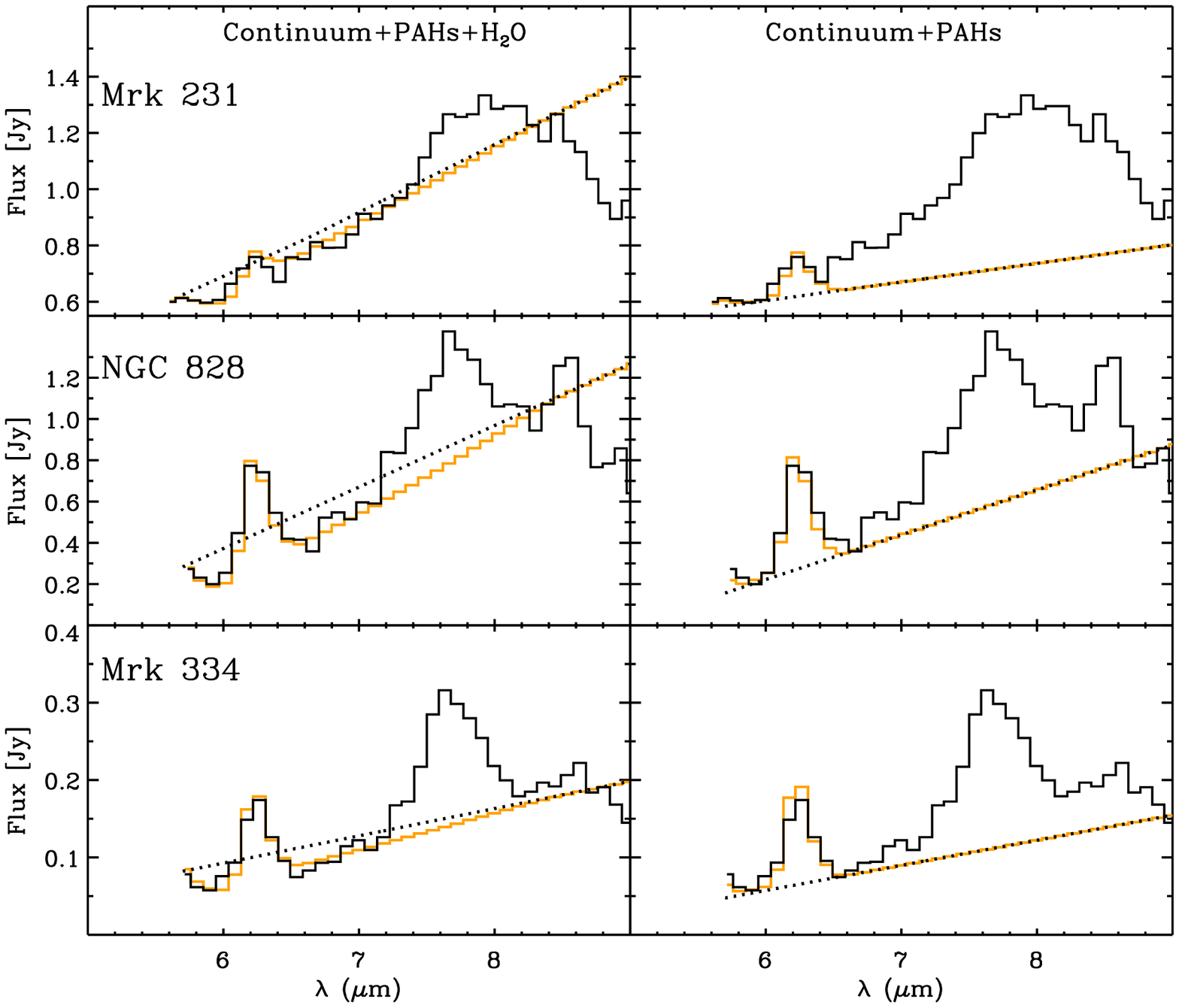,angle=90,width=400pt,height=300pt}
 \end{picture}
 \end{center}
\caption{Examples of 3 sources which show only a few pixels in the 
6.0 $\mu$m range and which might hinder the interpretation of H$_2$O 
ice being present. The left panel shows the best fits by a model 
containing both H$_2$O absorption and 5--6\,$\mu$m PAH emission. The 
right panel shows the best possible fits by a model which only contains 
5--6\,$\mu$m PAH emission features. The adopted continua are denoted by 
the dashed lines and the resulting fits are represented by the line 
shown in grey. Note that since our model does not take into account
the 7.7 and 8.6\,$\mu$m PAH features (see Sect.\,3.1), the quality 
of the fit can only be tested shortward of 7\,$\mu$m.}
\label{unsure_ice}
\end{figure*}

Fig.\,\ref{ice_fits} displays the model fits to all the sources, 
except for the three sources discussed below (Fig.\,\ref{unsure_ice}). 
The grey histogram represents the model and in all cases the model spectra 
match very well the observed data. For two sources, Arp\,220 and NGC\,4945, 
the \cam spectra have also been included (thin line). Reading from left to 
right, of the first 5 panels in Fig.\,\ref{ice_fits} a HAC absorption
feature has been included for four of the sources. This profile matches 
very well the strong absorption feature seen near 6.85\,$\mu$m in these
sources. In the case of I\,00188--0856, absorption between 6.4 and 
7.4\,$\mu$m is not consistent with the 6.85\,$\mu$m absorption profile
and therefore only the 6.0\,$\mu$m H$_2$O profile is modeled. We are
not aware of any other valid or likely candidate that can give rise to
the absorption observed in I\,00188--0856.
For the remainder of the sources, this feature has been
omitted from the model. A qualitative comparison of the panels of
Fig.\,\ref{ice_fits}, reveals that the observed features at
6.0\,$\mu$m and 6.2\,$\mu$m are successfully reproduced for all sources. 

For three of our ice galaxies (Mrk\,231, NGC\,828, Mrk\,334)
proving the presence of 6.0\,$\mu$m ice absorption -- the very key 
to membership of Classes 1--3  -- turned out to be non-trivial. 
For each of these galaxies (see Fig.\,\ref{unsure_ice}) the
number of pixels in the blue wing is small, as is the depth of the 
ice feature. In all cases continua exist which do not
require the presence of 6.0\,$\mu$m ice absorption. However, such
continua, which are defined here as the superposition of hot dust
plus PAH continua, must give a good fit in the 6--8\,$\mu$m range
(i.e., any emission above or below the continuum level should be
accounted for by known emission and absorption features) and the
continuum should join smoothly to the long wavelength ($\geq$12\,$\mu$m) 
data. In the case of Mrk\,231, there is a preference for the 
fit requiring the presence of water ice (Fig.\,\ref{unsure_ice}, 
left panel). This fit adopts a much more realistic continuum 
(cf. Fig.\,6) and it provides a more realistic value for the 
6.2\,$\mu$m/7.7\,$\mu$m PAH ratio than the fit not requiring the 
presence of water ice (Fig.\,\ref{unsure_ice}, right panel).  
Moreover, the fit without water ice would result in a very broad
7.7\,$\mu$m PAH feature which does not resemble PAH features in
Galactic sources. However, the total amount of ice absorption is 
difficult to quantify. A slightly lower but still permissible
continuum can reduce the ice optical depth by a factor 2. For the 
other two galaxies (NGC\,828, Mrk\,334) the fits with and without 
water ice are equally viable.

\subsection{Derived physical parameters}

Table\,\ref{optdepths} summarizes the physical parameters of the detected 
features. The range in optical depths of the \h2o absorption band is 
large, extending from small depths of $\tau_{\rm ice}$=0.13 
(I\,05189--2524) to very deep features with $\tau_{\rm ice}$=1.3 
(UGC\,5101). The optical depths of the \h2o band were converted to column 
densities by dividing the peak optical depth by the intrinsic peak 
strength (4.2 $\times \rm 10^{-20} \,cm^2$ molecules$^{-1}$; Hagen et al. 
\cite{Hagen}). The derived column densities are comparable 
to \h2o column densities determined for Galactic molecular clouds (Keane
et al. \cite{Keane}). All sources in Class 1, except for I\,00188--0856, 
also show evidence for additional absorption features centred at
6.85\,$\mu$m, 7.3\,$\mu$m, 7.67\,$\mu$m. The 6.85\,$\mu$m feature resides in
the long wavelength wing of the \h2o band, and hence the depth of this
feature is more accurately determined by dividing out the continuum
plus \h2o feature rather than just the continuum. For Class 1 galaxies,
except for I\,00183--7111, the depth of this feature is very strong and 
is greater than the corresponding \h2o optical depths. In Galactic 
dense molecular clouds, on the other hand, the water ice absorption
is always the stronger of the two (Keane et al. \cite{Keane}). The 
origin of the 6.85\,$\mu$m feature is unknown. The presence of a
7.3\,$\mu$m feature in some spectra suggests it is carried by HAC
residing in the diffuse ISM, very much like the Galactic Centre 
(Chiar et al. \cite{Chiar}). An absorption feature at 7.3\,$\mu$m 
is present in the spectra of I\,15250+3609 and NGC\,4418
(Spoon et al. \cite{Spoon01}). On the other hand, many
Galactic young stellar objects show a strong 6.85\,$\mu$m feature due in
part to an unidentified ice component (Keane et al. \cite{Keane}). 
Observations of the 3\,$\mu$m region could distinguish between these
two possibilities because HAC material presents a strong 3.4\,$\mu$m
absorption feature which is absent in Galactic YSOs (Pendleton et al. 
\cite{Pendleton94}; Pendleton \& Chiar \cite{Pendleton97}). A 3.4\,$\mu$m 
absorption feature observed towards a few extra-galactic sources has 
been tentatively attributed to HAC (Wright et al. \cite{Wright}).
Finally, one source, NGC\,4418, has an absorption feature at 7.67\,$\mu$m 
which has been attributed to methane ice (CH$_{4}$; Spoon et al. 
\cite{Spoon01}). 

Since the 6.2\,$\mu$m feature lies, in general, within the wing of the 
\h2o absorption band and hence slightly beneath the continuum, it 
is necessary to subtract a combination of the continuum plus \h2o from the 
observational data. After subtraction, the intensity (W\,cm$^{-2}$) of
the 6.2\,$\mu$m band is determined between 6.0\,$\mu$m and 6.55\,$\mu$m. 
The derived 6.2\,$\mu$m PAH 
intensities are listed in Table\,\ref{optdepths}. Note that for models
in which the water ice is mixed with or in front of the PAH emitting region, 
the intrinsic 6.2\,$\mu$m PAH intensities would be greater.

An estimate of the 9.7\,$\mu$m silicate optical depth ($\tau_{\rm sil}$)
for each of the sources is also given in Table\,\ref{optdepths}. In order
to be able to compute this quantity the 9.7\,$\mu$m continuum was 
interpolated from the 5--8\,$\mu$m continuum and the long wavelength
data ($>$12\,$\mu$m). For some sources the peak absorption of the silicate 
feature is saturated and the true depth of the band is thus uncertain.
In these cases fitting the wings of the feature can be attempted. Given 
the possibility of additional sources of emission along the line of sight, 
however, we prefer here to just state lower limits for $\tau_{\rm sil}$.
The column density of hydrogen is computed from the 9.7\,$\mu$m
silicate depth by assuming a Galactic conversion factor:-
{\it N}$_H$ = $\frac{{\it N}_H}{{\it A_V}}$ $\times$  $\frac{{\it
      A_V}}{{\tau_{sil}}}$ $\times$ $\tau_{sil}$ = 
${\rm{\tau_{sil} \times 3.5\times10^{22}\,cm^{-2}}}$; 
(Roche \& Aitken \cite{Roche}, Bohlin et al. \cite{Bohlin}). 

\begin{table*}[]
\caption{Observed physical parameters for the features residing in the 
3--8 $\mu$m spectral region.}
\begin{tabular}{lllcllllcll}
\hline
Target & Galaxy & cz & Class & H$_2$O ice  & H$_2$O ice  & PAH & HAC & HAC & Silicates & {\it N}(H$_2$O) \\
       & type   &    &       & 3.0\,$\mu$m & 6.0\,$\mu$m & 6.25\,$\mu$m & 6.85\,$\mu$m & 7.3\,$\mu$m & 9.7\,$\mu$m & \\
& & $[$km/s$]$ & & \hspace{.2cm} $\tau$ &\hspace{.2cm} $\tau$ &10$\rm^{-19}\,W cm^{-2}$ & \hspace{.1cm} $\tau$ & & \hspace{.1cm} $\tau$ & 10$\rm ^{18}\,cm^{-2}$ \\
\hline
Mrk\,334      & Sy2  &  6582& 3 &        &0.55&   2.0&    &        &    0.8 & 12.8 \\
NGC\,23       & SB   &  4566& 3 &        &0.94&   5.6&    &        &    1.0 & 22.0 \\
I\,00183--7111&Ulirg & 98032& 1 &        &0.20&   ---& 0.3&        & $>$1.9 &  4.6 \\
I\,00188--0856&Ulirg & 38550& 1 &        &0.55&   ---&    &        & $>$1.5 & 12.8 \\
NGC\,253      & SB   &   245&-- &0.25$^a$&    &      &    &        &        &      \\
I\,01173+1405 & SB   &  9362& 3 &        &0.70&   2.5&    &        &    1.3 & 16.3 \\
NGC\,828      & SB   &  5374& 3 &        &0.78&    11&    &        &    0.8 & 18.1 \\
I\,05189--2524& Sy2  & 12760& 2 &        &0.13&   1.1&    &        &    0.15&  3.0 \\
I\,06035--7102&Ulirg & 23823& 3 &        &0.70&   0.9&    &        &    1.5 & 16.3 \\
UGC\,5101     &Ulirg & 12000& 3 &        &1.30&   2.0&    &        & $>$1.5 & 30.2 \\
M\,82         & SB   &   203&-- &0.2$^a$ &    &      &    &        &        &      \\
NGC\,4418$^b$ & Sy?  &  2179& 1 &        &0.90&   ---& 1.1& $\surd$& $>$2.9 & 21.0 \\
Mrk\,231      &Ulirg & 12660& 2 &        &0.14&   1.5&    &        &    0.65&  3.3 \\
NGC\,4945     &SB/Sy2&   560& 3 &0.90$^c$&0.90&    47&    &        & $>$3.7 & 21.0 \\
Mrk\,273      &Ulirg & 11132& 3 &        &0.60&   1.5&    &        &    1.2 & 14.3 \\
I\,15250+3609 &Ulirg & 16000& 1 &        &1.20&   ---& 1.3& $\surd$& $>$3.3 & 28.0 \\
Arp\,220      &Ulirg &  5450& 2 &        &0.74&   2.3&    &        & $>$2.4 & 17.2 \\
I\,17208--0014&Ulirg & 12900& 3 &        &1.14&   3.3&    &        &    2.0 & 26.5 \\
I\,20551--4250&Ulirg & 12788& 2 &        &0.70&   1.4&    &        &    1.8 & 16.3 \\
I\,23128--5919&Ulirg & 13371& 3 &        &1.30&   2.5&    &        &    0.7 & 30.2 \\
\hline 
\multicolumn{11}{l}{$^a$ Sturm et al. (\cite{Sturm})}\\
\multicolumn{11}{l}{$^b$ Spoon et al. (\cite{Spoon01})}\\
\multicolumn{11}{l}{$^c$ Spoon et al. (\cite{Spoon00})}\\
\label{optdepths}
\end{tabular}
\end{table*}

\section{Discussion}

Within our sample of 103 galaxies with good S/N and sufficient 
wavelength coverage shortward of 6.0\,$\mu$m, we have found water
ice in up to 18 galaxies and upper limits of better than 
$\tau_{\rm ice}$=0.1--0.3 for the absence of water ice absorption 
in another 28 galaxies. Although a small sample, it is of interest
to correlate galaxy types with the mid-IR absorption/emission
characteristics. The results of our analysis are presented in
Table\,\ref{ice_stats}. 

Table\,\ref{ice_stats} suggests that water ice might be a common 
species in ULIRGs, since 12 out of 19 do show ice. Seyferts, on the 
other hand, seem to be ice-poor, with ice detected in only 2 out of 
62 galaxies. For starburst galaxies the numbers are less obvious, 
with 4 detections on a total of 21 galaxies.

\begin{figure}[]
 \begin{center}
  \psfig{figure=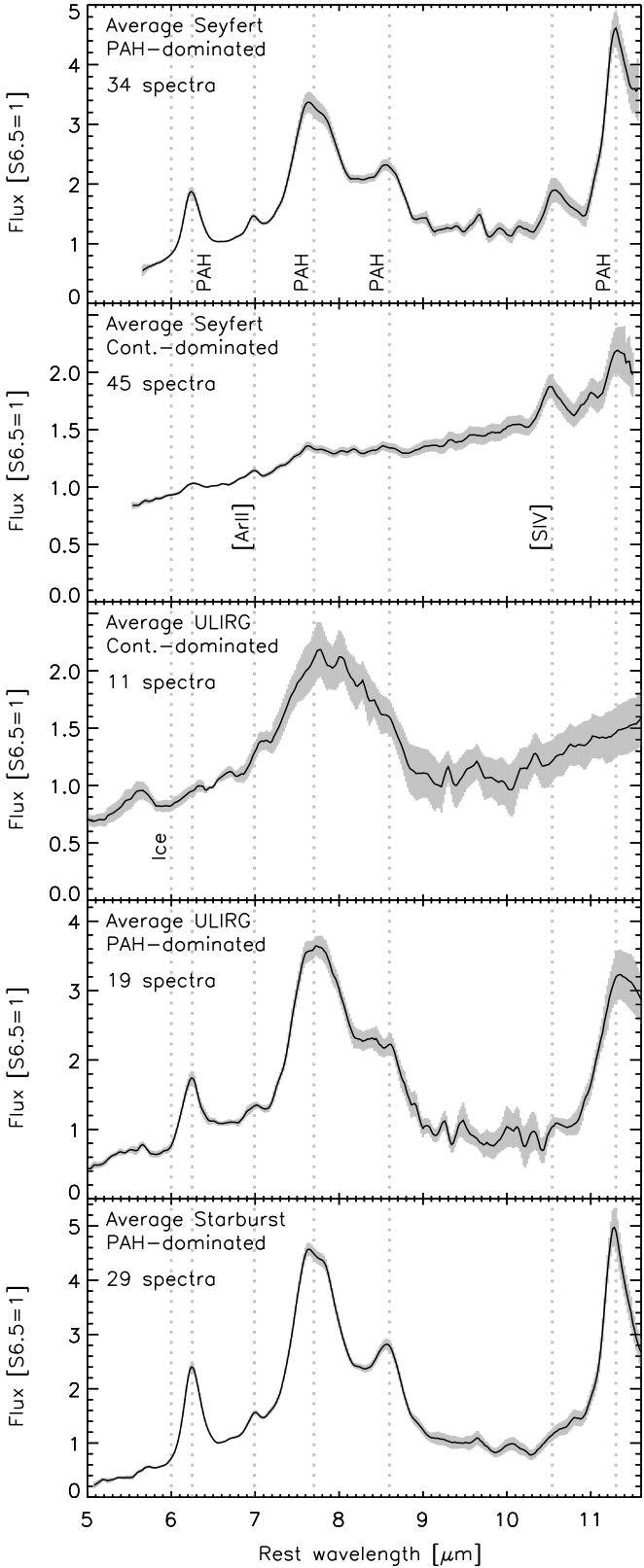,angle=0}
 \end{center}

\vspace*{-6mm}

\caption{Panels showing the average spectra of PAH-dominated 
(6.2\,$\mu$m-PAH L/C$>$0.15) and continuum dominated 
(6.2\,$\mu$m-PAH L/C$<$0.15) Seyferts, ULIRGs and starburst galaxies. 
Grey shaded areas indicate the standard deviations in the average 
spectra. Vertical dotted lines denote the central wavelengths of
well-known spectral features (see text).} 
\label{av_galaxies_panel}
\end{figure}

In order to investigate this issue further we have composed average
spectra for the three main galaxy types listed in Table\,\ref{ice_stats}.
Since the presence of 5--6\,$\mu$m PAHs in spectra complicates the
detection of the 6.0\,$\mu$m water ice feature, we have decided to 
generate separate average spectra for continuum-dominated Seyferts 
and ULIRGs and for PAH-dominated Seyferts and ULIRGs. We classify a 
spectrum as continuum-dominated when the 6.2\,$\mu$m PAH 
line-to-continuum ratio is less than 0.15 and PAH-dominated when 
the ratio exceeds this value. The resulting five average spectra are
shown in Fig.\,\ref{av_galaxies_panel} and are based on 138 out of 
250 galaxy spectra in our database. The galaxies included in the
averaging process were selected on the basis of their noise after 
scaling by their individual 6.5\,$\mu$m flux.

The top panel of Fig.\,\ref{av_galaxies_panel} shows the average
PAH-dominated Seyfert galaxy, composed of 34 galaxy spectra. 
In addition to the well-known PAH emission features, the spectrum
shows the 6.99\,$\mu$m $[$\ion{Ar}{ii}$]$ and 10.54\,$\mu$m
$[$\ion{S}{iv}$]$ forbidden lines, the latter of which usually is
strong in active galaxies. The spectrum bears no trace of
6.0\,$\mu$m water ice absorption. 
The second panel of Fig.\,\ref{av_galaxies_panel} depicts the 
average continuum-dominated Seyfert, composed of 45 galaxy 
spectra. The spectrum is dominated by the $[$\ion{S}{iv}$]$ line, 
with traces of $[$\ion{Ar}{ii}$]$, maybe of 7.65\,$\mu$m 
$[$\ion{Ne}{vi}$]$ and of weak PAH emission --- the latter in 
accordance with our selection criterion of 6.2\,$\mu$m-PAH L/C$<$0.15. 
The 9.7\,$\mu$m silicate absorption feature is strikingly 
absent, as is the 6.0\,$\mu$m water ice absorption feature.
The third panel of Fig.\,\ref{av_galaxies_panel} shows the average
continuum-dominated ULIRG, composed of 11 galaxy spectra, among which
are all Class 1 sources except for NGC\,4418, as well as the Class 2 
source Mrk\,231. The spectrum is dominated by a broad flux peak centered 
at $\sim$8\,$\mu$m. A 6.0\,$\mu$m water ice feature can be easily 
recognized, starting at $\sim$5.6\,$\mu$m and reaching maximum
depth at 5.9--6.0\,$\mu$m. The spectrum is noticeably different
from the other continuum-dominated spectrum, the Seyfert spectrum,
in the panel above.
The fourth panel shows the average PAH-dominated ULIRG, composed of
19 galaxy spectra. Among the selected sources are several Class 2 
and Class 3 sources. The spectrum shows strong resemblance to the average 
starburst galaxy (panel below; 29 spectra), with two exceptions: 
(1) the average PAH-dominated ULIRG shows a clear 6.0\,$\mu$m water ice 
feature, partially filled in by 6.2\,$\mu$m PAH emission, the average 
starburst galaxy does not; and (2) the average starburst spectrum 
shows a hint of 5--6\,$\mu$m PAH features (Fig.\,\ref{5_6_um_pahs}), 
where the average PAH-dominated ULIRG as well as the other three 
average spectra do not. The presence of the 5--6\,$\mu$m spectral 
structure near 6.0\,$\mu$m forces a higher upper limit to the amount 
of ice in the average starburst spectrum than in the two average 
Seyfert spectra.

The average spectra largely confirm our initial findings: ULIRGs
have on average more ice than Seyfert and starburst galaxies. 
The absence of contaminating PAH emission in the continuum-dominated 
Seyfert spectrum implies a tighter upper limit for the presence of 
water ice absorption than in the PAH-dominated Seyfert and starburst 
spectra.

Also of interest is the presence of a broad flux peak centered at 
$\sim$8\,$\mu$m in the continuum-dominated ULIRG 
(Fig.\,\ref{av_galaxies_panel}, middle panel). In the absence of 
strong 6.2\,$\mu$m PAH emission 
(L/C$<$0.15), the 8\,$\mu$m flux peak cannot be attributed to strong 
and exceptionally broad PAH emission. We will come back to this issue 
later this Section.

\begin{table}[t]
\caption{Break-down of ice galaxies over the three main galaxy types 
in our sample of 103 ISO galaxies having good S/N spectra and sufficient 
wavelength coverage blueward of 6\,$\mu$m.}
\begin{tabular}{lccc}
\hline
Gal. type & PHT-S & CAM-CVF & total \\
\hline
Seyfert   &  2/55 & 0/7  &  2/62 \\ 
ULIRG     &  9/12 & 3/7  & 12/19 \\
Starburst &  3/12 & 1/9  &  4/21 \\
Other     &  0/1  & 0/0  &  0/1  \\
\hline
\label{ice_stats}
\end{tabular}
\end{table}

The galaxies in our sample have turbulent and hostile environments
(i.e., the diffuse interstellar medium or the immediate surroundings
of the central starburst or AGN) which significantly process and often
destroy the molecular material that reside in or close to these
regions. However, within our ice-galaxy sample, volatile material
(which is easily destroyed) is clearly present. This suggests that some
shielding from harsh environments must occur. In the diffuse medium
supernova shock waves are the predominant destruction mechanism for
interstellar dust (Jones et al. \cite{Jones}). Refractory grains and most
likely PAHs are typically destroyed within 10$^{8}$\,yr, whereas ice
grains of 1000\,${\rm\AA}$ can only survive for $\sim$\,10$^{6}$\,yr. These
lifetimes increase significantly in dense clouds where the molecules
are protected from processes such as sputtering by shocks. On the
other hand, molecular material in an AGN torus (or in nearby dusty
clouds) is exposed to the intense hard X-rays originating from the
central engine. Due to their extreme energies, hard X-ray photons are
able to penetrate column densities of $\sim$\,10$^{25}\rm\,cm^{-2}$
(e.g. NGC\,4945: Iwasawa et al. \cite{Iwasawa}). Hence, under such 
conditions, ice mantles on the dust grains
would be unable to survive. PAHs, on the other hand, might be able to
survive for lifetimes of $>$\,10$^{4}$\,yr if the column density is
greater than $\sim$\,10$^{23}$cm$^{-2}$ (Voit \cite{Voit}) and hence it 
is more likely to see PAHs as opposed to ices in such environments. 
Nevertheless, the harsh
conditions in AGN toroids may well influence the composition of the
emitting PAH family.  Further modelling and laboratory experiments are
required to determine to what extent PAHs and ices are processed by the
hard radiation present in AGNs.

ULIRGs are merger systems in which the tidal forces have allowed 
molecular material from the galaxy disks to accumulate in the nuclei 
of the system (e.g. Solomon et al. \cite{Solomon}; Downes \& Solomon
\cite{Downes}; Sakamoto et al. \cite{Sakamoto}; Tacconi et al.
\cite{Tacconi}). The presence of water ice in ULIRGs thus comes as no
surprise. In the many dense shielded cloud cores, that are 
likely to be present, water ice can survive until these clouds are 
dispersed by starformation. Clearly, the ``icy'' ULIRGs still have 
ample cold dense molecular material to continue star formation for 
some time.
In Galactic dense molecular clouds the 6.0\,$\mu$m water ice 
absorption feature is seen against the mid-IR continuum of deeply 
embedded protostars, like W\,33A and NGC\,7538\,IRS9. 
In external galaxies, the background continuum source might also 
be provided by hot dust associated with the AGN torus or with 
starburst activity. The similarity of the average ULIRG spectrum 
(Fig.\,1 of Lutz et al. \cite{Lutz98}) to Class 3 spectra with strong 
PAHs, suggest that ice absorption in ULIRGs is preferentially linked 
to a scaled up star formation process. 
The lower luminosity example of NGC\,4418 cautions, however, that 
this link cannot be generalized to all individual objects: ice 
absorption may also be seen against AGN.

Little can be derived from our data on the ice absorption properties 
of lower luminosity starbursts. The very strong PAH emission features 
prevent the detection of moderately strong 6.0\,$\mu$m features. In fact, 
given the lower obscuration of starbursts compared to ULIRGs or to 
Galactic protostars, weaker ice features in starbursts are expected 
for similar properties of the obscuring matter. The examples of M\,82 
and NGC\,253 (Sturm et al. \cite{Sturm}) suggest that the 3\,$\mu$m 
region may be better suited to quantify such weaker ice absorptions, 
with the caveat for ground-based observations of the 3\,$\mu$m region 
being a difficult atmospheric window. Ice is seen in these starbusts at 
3\,$\mu$m but not detectable in the heavily structured 6\,$\mu$m 
spectrum.

The conditions for the existence of water ice near Seyfert nuclei seem 
to be generally poor. If most of the obscuration occurs in a dense torus 
that is surrounding the nuclear region and heated by the X-rays from 
the central engine, ices are unlikely to survive. Indeed, 3\,$\mu$m 
spectra of AGN (Bridger et al. \cite{Bridger}; Wright et al. 
\cite{Wright}; Imanishi \cite{Imanishi00}; Sturm et al. \cite{Sturm}) 
tend to show 3.4\,$\mu$m hydrocarbon absorptions but no ices.

The low number of Seyfert and starburst ice-galaxies, as inferred 
from the mid-IR spectra, might indicate that dense shielded molecular 
cloud cores are far less abundant in these systems than in ULIRGs. 
Likely, most of the absorption in Seyferts originates in the dense 
toroid surrounding the nuclear region, 
which is thoroughly energetically or thermally processed by the 
X-rays of the internal monster and therefore contains no ices.

\begin{table}[t]
\caption{Overview of features observed in Class 4--6 galaxies
shown in Figs.\,\ref{5_6_um_pahs}, 
\ref{sil_no_ice_panel}\,\&\,\ref{no_sil_no_ice_panel},
all displaying no sign for 6\,$\mu$m water ice absorption at the 
$\tau$(6\,$\mu$m\,ice)=0.1--0.3 level.}
\begin{tabular}{lllcc}
\hline
Target & Galaxy & cz         & 5\,$\mu$m & 9.7\,$\mu$m \\
       & type   &            & PAH       & silicate    \\
       &        & $[$km/s$]$ & emission  & absorption  \\
\hline
NGC\,253       & SB    &  245 & ++   & ?  \\
NGC\,520       & SB    & 2281 & ++   & +  \\
NGC\,613       & SB    & 1475 & --   & ?  \\
NGC\,1068      & Sy2   & 1148 & --   & +  \\
NGC\,1365      & Sy1.8 & 1636 & ?    & ?  \\
I\,04385--0828 & Sy2   & 4527 &      & ++ \\
NGC\,1808      & Sy2   & 1000 & ++   & ?  \\
MGC\,8--11--11 & Sy1.5 & 6141 &      & -- \\
M\,82          & SB    &  203 & ++   & ?  \\
M\,83          & SB    &  516 & ++   & ?  \\
Mrk\,279       & Sy1.5 & 8814 &      & -- \\
Circinus       & Sy2   & 439  & ?    & ++ \\
PKS\,2048--57  & Sy2   & 3402 &      & +  \\
\hline
\multicolumn{5}{l}{++ = strongly present} \\
\multicolumn{5}{l}{+  = present} \\
\multicolumn{5}{l}{?  = maybe present} \\
\multicolumn{5}{l}{-- = absent} \\
\label{iceless_stats}
\end{tabular}
\end{table}

As a corollary and as illustrated by Fig.\,\ref{sil_no_ice_panel} the 
presence of silicates does not automatically imply the presence of water 
ice. If the environment is not sufficiently shielded or is too warm, 
water ice cannot exist. For the five galaxies displayed (one starburst 
galaxy and four Seyferts) this might be the case.

Conversely, does water ice correlate with the presence of shielding dust
(e.g., silicates)\,? Except for I\,05189--2524, all our ice-galaxies 
show a strong 9.7\,$\mu$m silicate absorption feature (see 
Table\,\ref{optdepths}), with $\tau_{\rm sil}>$1.5 for most sources. The 
ratio $\tau_{\rm ice}$/$\tau_{\rm sil}$ spans a range of $<$0.10 for
I\,00183--7111 to 1.8 for I\,23128--5919, with a mean value of 0.67 for
the galaxies with a non-saturated silicate feature. In contrast, all
five Class 1 galaxies have $\tau_{\rm ice}$/$\tau_{\rm sil}<$0.37. 
This suggests that silicate and water ice optical depths are not 
correlated on the galaxy-wide scale that we probe with our ISO
observations. We note that low mass protostars in individual clouds 
(e.g. Taurus) do show a good correlation between ice optical depth
and the strength of the silicate feature (Whittet et al. 
\cite{Whittet83}, \cite{Whittet88}). However, for massive luminous 
protostars in different clouds, no such correlation exists, largely due 
to the extensive thermal processing of the environment by the newly 
formed star (cf. Tielens \& Whittet \cite{Tielens}). Obviously, the 
abundance of water ice is controlled by special conditions, which in our 
sample are apparently only met in some of the regions within our Class 1 
ice galaxies.

Though originally published as a unique source, it is apparent from our 
study that some characteristics of NGC\,4418 are seen in a larger number 
of galaxies. Mid-IR absorptions of water ice are detected in about one 
sixth of the galaxies for which ISO spectra of adequate S/N are available, 
and perhaps the majority of ULIRGs. These detections range from spectra 
dominated by an absorbed continuum (Class 1) over spectra mainly showing 
absorption signatures but also clear PAH emission (Class 2) to spectra 
that resemble normal PAH emission spectra subjected to additional dust 
and ice obscuration (Class 3). 

\begin{figure}[]
 \begin{center}
  \psfig{figure=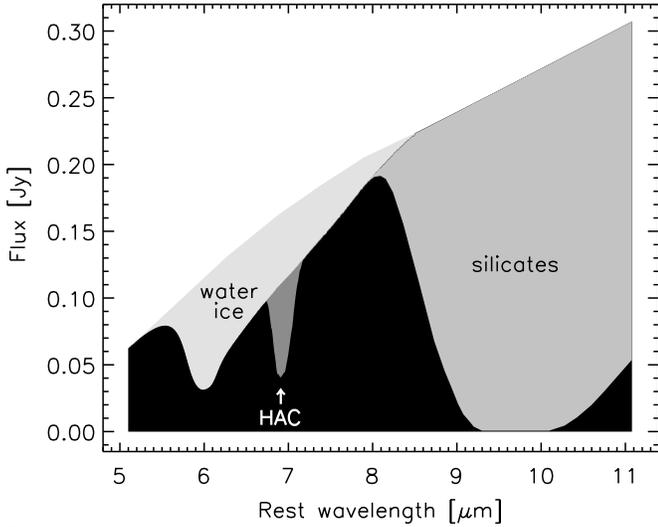,angle=0}
 \end{center}

\vspace*{-7mm}

\caption{Schematic view of the impact of dust and ice absorption on 
a mid-IR continuum spectrum. All shaded areas combined
constitute the adopted local mid-IR continuum. The black area is
all that is left of the local continuum after ices and silicates
absorb substantial portions of the 5--12\,$\mu$m local continuum.
Note the presence of an emission-like broad feature near 8\,$\mu$m in
the resulting observed spectrum, which at first glance or at poor S/N 
may be mistaken for 7.7\,$\mu$m PAH emission.}
\label{8um_bump}
\end{figure}

Short of full spectral modelling, empirical criteria help to assess the 
relative importance of emission and absorption in shaping these spectra. 
First and most important is the distinction between a 6.2\,$\mu$m PAH 
emission and the 6.5--6.7\,$\mu$m pseudo maximum in an absorption dominated 
spectrum (Spoon et al. \cite{Spoon01}). Second, there are differences in 
the shape of the maximum in the 8\,$\mu$m region. PAH dominated spectra 
exhibit relatively narrow 7.7 and 8.6\,$\mu$m emissions (e.g. Helou et al.
\cite{Helou}, Rigopoulou et al. \cite{Rigopoulou}). Also, they show a very 
steep rise between 7.2 and 7.7\,$\mu$m. Obscuration of such a spectrum 
(assumed to be intrinsically unchanged) affects mainly the relative 
strengths of such features but keeps the 7.7\,$\mu$m peak narrow 
($\approx$0.6$\mu$m FHWM, e.g. Fig.\,\ref{extinctionlaws}; Fig.\,6 of 
Rigopoulou et al. \cite{Rigopoulou}) and its ascent steep. In contrast, 
absorption dominated spectra tend to show a wider, less well defined peak 
around 8\,$\mu$m (e.g. Fig.\,\ref{8um_bump}), with significant variation
depending on continuum shape and optical depth of the various features.

\begin{figure}[]
 \begin{center}
  \psfig{figure=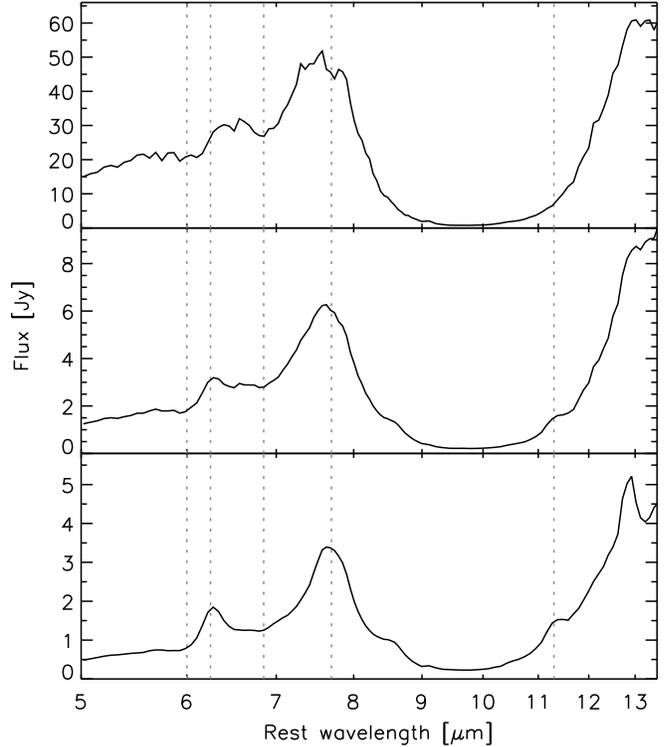,angle=0}
 \end{center}
\caption{ISO-CAM-CVF spectra of three adjacent positions in the W\,3 
molecular cloud (D.\,Cesarsky, priv. comm.), showing a gradual
transition between an obscured PAH-dominated mid-IR spectrum (bottom) 
and a strongly absorbed hot dust continuum (top). The vertical dotted 
lines show the positions of 6.0\,$\mu$m water ice, 6.25\,$\mu$m PAH, 
6.85\,$\mu$m HAC and 7.7\,$\mu$m \& 11.3\,$\mu$m PAH.}
\label{w3_panels}
\end{figure}

The change from an absorbed continuum to an absorbed PAH spectrum is 
illustrated further in spectra of three adjacent positions in the Galactic
star forming complex W\,3 (Fig.\,\ref{w3_panels} , D.\,Cesarsky, priv. 
comm.). The top panel towards the infrared source 
is dominated by a continuum absorbed by ice, HAC, and silicates with a 
broadish 8\,$\mu$m bump. The middle panel is an intermediate case and the 
bottom shows an obscured PAH spectrum. These spectra illustrate that the 
qualitative features discussed above and outlined in the schematic views 
of Fig.\,\ref{extinctionlaws} and Fig.\,\ref{8um_bump} can indeed 
originate from radiation transfer through a dense and dusty medium, and 
that star formation in our own galaxy can reproduce the full range of 
phenomena.

Differences in the shape of the 8\,$\mu$m maximum may help in 
understanding the vast majority of sources that are not pure absorbed 
continua. Is the frequently observed weakness of the 6.2\,$\mu$m feature 
(compared to the 7.7/8\,$\mu$m maximum) due to obscuration of a 
PAH-dominated spectrum or due to superposition of a little obscured PAH 
spectrum and a strongly obscured continuum? The intrinsic 6.2/7.7 PAH 
ratio is quite stable for many Galactic sources (Peeters et al. 
\cite{Peeters}) as well as the ISM in normal and starburst galaxies 
(Helou et al. \cite{Helou}; Rigopoulou et al. \cite{Rigopoulou}, our 
Fig.\,\ref{5_6_um_pahs}). Lutz et al. (\cite{Lutz98}) and Rigopoulou 
et al. (\cite{Rigopoulou}) argued in favour of obscuration as cause of 
the 6.2 weakness, using an observed correlation between 6.2/7.7 feature 
ratio and extinction to the starburst region, as measured from independent 
mid-IR emission line data (Genzel et al. \cite{Genzel}). On the other 
hand, suggestions have been made of an effectively unobscured 'surface 
layer' producing most of the PAH emission, CII line emission, and submm 
continuum of ULIRGs (e.g. Fischer et al. \cite{Fischer}, Haas et al. 
\cite{Haas}). Such an approach tackles problems like the observed 
$[$\ion{C}{ii}$]$ deficit of ULIRGs at the expense of breaking the link 
between PAHs and the star formation observed in the mid-infrared fine 
structure lines, and of introducing an unknown fully obscured component. 

Water feature optical depths of $\tau_{\rm ice}\sim$1 (Table\,\ref{optdepths}) 
are sufficient to modify the ratio of the 6.2 and 7.7\,$\mu$m PAH features 
by a factor $\sim$2. While this is enough to explain the weakness of 
the 6.2\,$\mu$m feature in many objects by obscuration, both the uncertainty 
in the optical depth values and the lack of knowledge to which extent the 
ice features apply to the PAHs and/or an underlying continuum make 
conclusions for individual objects very difficult. Here, use of the shape 
of the feature near 7.7/8\,$\mu$m can help with breaking the degeneracy. As 
long as the intrinsic shapes of the PAH features are assumed to remain 
constant, a wider than usual peak near 8\,$\mu$m argues for a considerable 
contribution of a heavily absorbed continuum. Our Class 1 and 2 spectra 
(Fig.\,\ref{class1} and \ref{class2}) suggest that this is the case for 
a number of galaxies. In a future paper (Spoon et al. in prep.) we will 
derive quantitative fits to the high quality Class 2 spectrum of Arp\,220, 
arguing for a best fit with a considerable contribution by an absorbed 
continuum. The existence of Class 1 and Class 2 sources implies that 
estimates of the PAH contribution on the basis of a line-to-continuum ratio
(Genzel et al. \cite{Genzel}) or of fits with extinction laws that do 
not include ices (Tran et al. \cite{Tran}) will overestimate the importance 
of PAHs, more noticeably for the simple line-to-continuum ratio. This 
effect is highly relevant for some of the Class 1 and 2 spectra, but less 
so for Class 3 or the ULIRGs in general. 
High S/N spectra, extended wavelength coverage, and fits of the entire 
wavelength range with PAHs and continua obscured by dust and ices are 
needed for quantitative progress.

The issue of broad 8\,$\mu$m features is complicated further by the 
presence of yet another category of "8\,$\mu$m maxima" in addition to 
PAHs and absorbed spectra: Several luminous AGN-like ULIRGs (Tran et al. 
\cite{Tran}; Taniguchi et al. \cite{Taniguchi}) show broad 8\,$\mu$m 
maxima on top of a smooth mid-infrared continuum, with PAH absent or 
very weak. Clear examples for this include I\,09104+4109, I\,00275--2859, 
I\,22192--3211, and I\,23529--2119. The key difference to our Class 1 
and Class 2 spectra with broad 8\,$\mu$m maxima is that these objects do 
not show the very deep silicate feature of the absorbed Class 1 and 2 
spectra. Tran et al. (\cite{Tran}) discussed these features in terms of 
self-absorbed silicate emission or a modified PAH origin, without definite 
conclusion. Recent surveys of galactic PAHs (Peeters et al. 
\cite{Peeters}; Verstraete et al. \cite{Verstraete}) fail to 
observe similar profiles even under unusual conditions. The nature of 
these features remains uncertain.

It is interesting to speculate in analogy to Galactic sources that our 
classification might reflect an evolutionary sequence. If an evolved 
starburst is represented by a a PAH spectrum, while strong mid-infrared 
continua are typical for the $\ion{H}{ii}$-region continuum of intense 
compact starbursts (Laurent et al. \cite{Laurent}) or AGN, then Class 1 
may trace the deeply embedded beginnings of star formation (or AGN 
activity), while the latter classes reflect more advanced and less 
enshrouded stages.

Finally, a point of caution. Since the 3.0\,$\mu$m water ice absorption 
extends over a difficult part of the L band, its definition is not easy in 
ground based spectra like those of Imanishi et al. (\cite{Imanishi01}). Its 
short end is outside the L band and its long end coincides with the 
3.3\,$\mu$m PAH emission and 3.4\,$\mu$m hydrocarbon absorption. A gradually 
increasing continuum from 3.15 to 4.0\,$\mu$m is however visible in 
their spectrum of UGC\,5101. Depending on the location along the line of 
sight, the water ice absorption will strongly affect measured quantities 
like the 3.3\,$\mu$m continuum, the 3.3\,$\mu$m PAH flux, its equivalent 
width and the 3.3\,$\mu$m-PAH-to-far-IR flux ratio. Hence, in galaxies with 
water ice absorption (like UGC\,5101, $\tau_{\rm ice}$=1.3), both 
uncorrected 3.3\,$\mu$m and 6.2\,$\mu$m PAH fluxes will provide 
underestimates when used as star formation indicators. For the 3.3\,$\mu$m 
PAH, aperture losses of present instruments are an additional reason for 
underestimates as in the case of UGC\,5101 where the north-south oriented 
1.2$\arcsec$ slit of Imanishi et al. (\cite{Imanishi01}) is perpendicular 
to the orientation of the $>$2$\arcsec$ star forming region (Genzel et al. 
\cite{Genzel}).

\section{Conclusions}

Following the discovery of water ice in NGC\,4418 (Spoon et al. \cite{Spoon01}), 
we have searched our sample of ISO galaxy spectra for galaxies showing 
similar signs of 6.0\,$\mu$m water ice absorption. We have found in total 
18 galaxies, which we grouped into three classes. The classification is 
based on the presence of ice absorption, 6.2\,$\mu$m PAH emission, and the 
nature of the 7.7--8\,$\mu$m feature: PAH emission or absorbed mid-IR 
continuum, or a combination of the two. We also looked for galaxies showing
no signs of water ice absorption. This sample contains 28 galaxies with water 
ice upper limits ranging from $\tau_{\rm ice}$=0.1--0.3, depending on the 
S/N and the complexity of the 5--7\,$\mu$m spectrum. We classified these
galaxies in another three classes (Class 4--6), depending on the presence 
of 9.7\,$\mu$m silicate absorption and 5--6\,$\mu$m PAH emission.

We have modeled the complicated interplay of 6.0\,$\mu$m water ice 
absorption and 6.2\,$\mu$m PAH emission, which in ISO-PHT-S spectra 
takes place at the blue end of the PHT-SL range and, hence, only can be 
recognized in galaxies with a redshift in excess of 3000 km/s. For spectra
obtained with the ISO-CAM-CVF instrument no such limitations apply. Our 
modeling supports the presence of ice in the Class 1--3 sources, except in 
a few cases where ice is not strictly required to obtain an acceptable fit
to the observed spectrum.

Based on a subsample of 103 good S/N ISO galaxy spectra with sufficient
wavelength coverage blueward of 6\,$\mu$m, a substantial fraction 
of ULIRGs (12 out of 19) contain detectable amounts of water ice. On
the other hand, the majority of Seyfert (2 out of 62) and starbursts 
galaxies (4 out of 21) probably do not. These results are confirmed
by the spectral structure seen in our average spectra of Seyfert, ULIRG 
and starburst galaxies: water ice absorption is only obvious in the 
average ULIRG. 

Class 1\,\&\,2 ice galaxies are dominated by a broad feature near 8\,$\mu$m 
which indicates a strong contribution by a dust and ice-absorbed continuum, 
similar to that seen in NGC\,4418. These observation stress the need for high
S/N data and refined diagnostic methods, to properly discriminate spectra 
dominated by PAH emission and spectra dominated by heavy obscuration.

The interplay of the broad 8\,$\mu$m feature and PAH emission, as seen in 
our ice galaxies, shows strong similarities with features seen in Galactic 
star forming clouds. This leads us to believe that our classification 
of ice galaxy spectra in three classes might reflect an evolutionary 
sequence from strongly obscured beginnings of star formation (and AGN 
activity) to a less enshrouded stage of advanced star formation (and AGN 
activity), as the PAHs get stronger and the broad 8\,$\mu$m feature 
weakens.

\acknowledgements
The authors wish to thank Diego Cesarsky, Els Peeters and Eckhard Sturm 
for discussions.
This research has made use of the NASA/IPAC Extragalactic Database (NED) 
which is operated by the Jet Propulsion Laboratory, California Institute 
of Technology, under contract with the National Aeronautics and Space
Administration.

\end{document}